\documentclass[reprint, superscriptaddress, amssymb, aps, prx]{revtex4-2}

\usepackage[usenames,dvipsnames]{xcolor} 
\usepackage{graphicx}
\usepackage{mathtools}
\usepackage{dcolumn}
\usepackage{bm}
\usepackage[colorlinks=true,citecolor=blue,linkcolor=blue]{hyperref}
\usepackage[caption=false]{subfig}
\usepackage{bbm}
\usepackage{braket}
\usepackage{comment}
\usepackage{hyperref}
\usepackage{upgreek}
\usepackage{esint}
\usepackage{soul}
\usepackage[normalem]{ ulem } 
\usepackage{floatrow}

\graphicspath{ {figures/} }

\makeatletter
\def\p@subsection{}
\makeatother

\renewcommand{\vec}[1]{\boldsymbol{\mathbf{#1}}}

\global\long\def\sgn{\operatorname{sgn}}

\newcommand{\normalorder}[1]{:\mathrel{\mspace{2mu}#1\mspace{2mu}}:}

\definecolor{prio-color}{RGB}{255,0,0}

\newcommand{\EqualContribution}{N.~S.\ and B.~A.~K.\  contributed equally to this work}

\begin{document}

\title{Superconductivity and fermionic dissipation in quantum Hall edges}

\author{Noam Schiller}
\thanks{\EqualContribution}
\affiliation{\mbox{Department of Condensed Matter Physics, Weizmann Institute of Science, Rehovot 76100, Israel}}
\author{Barak A. Katzir}
\thanks{\EqualContribution}
\affiliation{\mbox{Physics Department, Technion, 320003 Haifa, Israel}}
\author{Ady Stern}
\affiliation{\mbox{Department of Condensed Matter Physics, Weizmann Institute of Science, Rehovot 76100, Israel}}
\author{Erez Berg}
\affiliation{\mbox{Department of Condensed Matter Physics, Weizmann Institute of Science, Rehovot 76100, Israel}}
\author{Netanel H. Lindner}
\affiliation{\mbox{Physics Department, Technion, 320003 Haifa, Israel}}
\author{Yuval Oreg}
\affiliation{\mbox{Department of Condensed Matter Physics, Weizmann Institute of Science, Rehovot 76100, Israel}}

\date{\today}

\begin{abstract}
    Proximity-induced superconductivity in fractional quantum Hall edges is a prerequisite to proposed realizations of parafermion zero-modes. A recent experimental work [G\"{u}l et al., Phys. Rev. X 12, 021057 (2022)] provided evidence for such coupling, in the form of a crossed Andreev reflection signal, in which electrons enter a superconductor from one chiral mode and are reflected as holes to another, counter-propagating chiral mode. Remarkably, while the probability for crossed Andreev reflection was small, it was stronger for $\nu=1/3$ fractional quantum Hall edges than for integer ones. We theoretically explain these findings, including the relative strengths of the signals in the two cases and their qualitatively different temperature dependencies. An essential part of our model is the coupling of the edge modes to normal states in the cores of Abrikosov vortices induced by the magnetic field, which provide a fermionic bath. We find that the stronger crossed Andreev reflection in the fractional case originates from the suppression of electronic tunneling between the fermionic bath and the fractional quantum Hall edges. Our theory shows that the mere observation of crossed Andreev reflection signal does not necessarily imply the presence of localized parafermion zero-modes, and suggests ways to identify their presence from the behavior of this signal in the low temperature regime. 
\end{abstract}

\maketitle 
    

\noindent{\it Introduction}---
Topological quantum computation (TQC) benefits from resilience to errors arising from local noise and decoherence processes \cite{Kitaev_2003, Nayak_2008}. In particular, such protection is obtained by encoding the quantum data in many body systems which harbor a non-Abelian phase of matter. These phases are characterized by having an energy gap and non-trivial ground state degeneracy in the presence of specific quasiparticles or defects \cite{Stern_2010}.

The most well-studied non-Abelian phases are those supporting Majorana zero-modes (MZMs). Notable examples of systems realizing such phases are the Moore-Read fractional quantum Hall state~\cite{Moore_1991, Stern_2008}, $p+ip$ superconductors~\cite{Read_2000, Ivanov_2001} and arrays of $p$-wave superconducting wires \cite{Kitaev_2001, Oreg_2010, Lutchyn_2010, Cook_2011}. While currently being the most experimentally accessible non-Abelian systems, MZMs do not admit universal TQC~\footnote{The set of quantum gates associated with both TQC based on MZMs and PZMs is limited to qubit and qudit Clifford gates respectively and as such is non-universal}. Beyond the MZM paradigm, phases admitting parafermion zero-modes (PZMs) support a richer set (yet not universal) of topologically protected quantum gates induced by braiding \cite{Note1}. Furthermore, systems supporting universal TQC can be realized by utilizing an array of PZMs \cite{Mong_2014}.

\floatsetup[figure]{style=plain,subcapbesideposition=top}
\begin{figure}[t!]
    \centering
    \sidesubfloat[]{
        \includegraphics[scale=0.92]
            {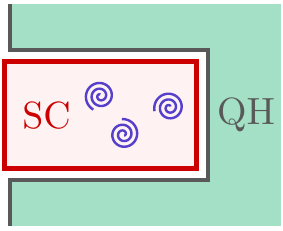}
        \label{subfig:finger_illus}
    }
    \ 
    \sidesubfloat[]{
        \includegraphics[scale=0.92]
            {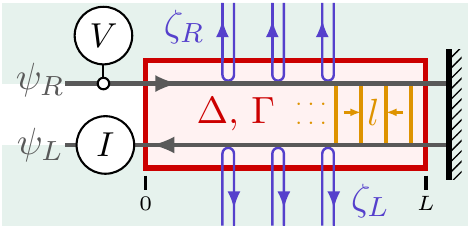}
        \label{subfig:model_illus}
    }
    \\[\baselineskip]
    \sidesubfloat[]{
        \includegraphics[width=0.8\columnwidth]
            {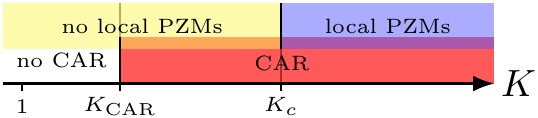}
        \label{subfig:phase_dia}
    }
    \caption{
        \protect\subref{subfig:finger_illus} The physical system: a quantum Hall (QH) droplet, with its edge proximity coupled to a superconductor (SC) in a ``finger'' shape. Due to the strong magnetic field Abrikosov vortices are present in the SC finger and act, due to their normal core, as a metallic bath. \protect\subref{subfig:model_illus} The model, Eq.~\eqref{eq:Hamiltonian}: the boundary of a QH droplet at filling factor $\nu=1/m$ (for odd integer $m$) is modeled by two edge modes, $\psi_{R/L}(x)$, counter-propagating along a grounded SC finger of length $L$. The right-moving mode arrives at the finger region $x=0$ at bias $V$, and is totally back-scattered at $x=L$ into a left-moving edge mode that leaves the finger region  at $x=0$. The counter-propagating edges are proximity coupled to the SC finger, and are subject to back-scattering and Andreev reflection between the edges. We denote the tunneling amplitudes of these two processes by $\Gamma$ and $\Delta$ respectively. The dissipative vortices are modeled as one dimensional metallic modes $\zeta_{R/L}$ coupled at each point along the finger to the QH edges. \protect\subref{subfig:phase_dia} A phase diagram of a fractional model with $m>2$ as a function of the Luttinger parameter $K$. Above $K>K_{\text{CAR}}$ there is the onset of CAR correlations. At $K=K_c$ we find a phase transition to a local PZM-harboring phase.
    }
    \label{fig:system_illus}
\end{figure}


A promising method to realize PZMs utilizes proximity coupling between a superconductor (SC) and counter-propagating fractional quantum Hall (FQH) edge modes~\cite{lindner_fractionalizing_2012, clarke_exotic_2013, Cheng_2012, Vaezi_2013, alicea_topological_2016} (other routes towards realizing PZMs~\cite{Zhang_2014, Orth_2015, Fleckenstein_2019, Barkeshli_2013_twist, barkeshli_experimental_2014} and various experimental signatures~\cite{schiller_predicted_2020, michelsen_current_2020,snizhko_measurement_2018,snizhko_parafermionic_2018,nielsen_readout_2021} have been proposed as well). Importantly, to sustain PZMs suitable for TQC, this coupling must induce an energy gap. 
This requires overcoming two major obstacles. First, the gap can be impeded by repulsive electron-electron interactions within the edges. This obstacle can be overcome for judicious choices of physical parameters \cite{katzir_superconducting_2020}. 
The second challenge, and the focus of this work, is that the high magnetic fields required to sustain the FQH state clash with superconductivity. This leads to a proliferation of in-gap states, such as those residing in the cores of Abrikosov vortices in the superconductor supplying the proximity coupling  \cite{gennes_superconductivity_2019, tinkham_introduction_2004,caroli_bound_1964}. By electron-tunneling between these states and the edge modes, current can be conveyed out of the system through a grounded reservoir connected to the superconductor, serving as a source of fermionic dissipation. 

Despite these challenges, systems of SCs proximity coupled to QH edge modes have been subject to much interest. This includes theoretical approaches to predict transport measurements of these systems \cite{manesco_mechanisms_2022,michelsen_supercurrent-enabled_2022,galambos_crossed_2022,kurilovich_disorder_2022}, and experimental works showing promising results in graphene \cite{Ronen_2020, lee_inducing_2017, hatefipour_induced_2021, amet_supercurrent_2016,zhao_interference_2019}. In particular, Ref.~\cite{Ronen_2020} successfully confirmed proximity coupling of both integer quantum Hall (IQH) and FQH states via the observation of crossed Andreev reflection~(CAR) across the superconductor. Notably (see Fig.~5f in~Ref.~\cite{Ronen_2020}), the probability of CAR was larger for Laughlin FQH states than IQH; however, in both cases, this probability was significantly smaller than 1, hinting that fermionic dissipation plays a major role.
In the IQH case, the CAR probability was largely temperature independent. In contrast, the proximitized Laughlin FQH edges displayed a CAR signal that grows with decreasing temperature, raising the hope that the system approaches the local PZM-harboring phase in the zero temperature limit.

Motivated by these results, in this paper we introduce a model to treat the effects of vortices on electric transport QH-SC hybrid systems. Our model includes three types of processes: dissipation via single-electron tunneling into vortex core states; back-scattering between edges; and crossed Andreev reflection via the superconductor. We note that while dissipation has been studied in Luttinger liquids in various contexts \cite{altland_intermediate_2015,cazalilla_dissipation-driven_2006,friedman_dissipative_2019,ristivojevic_transport_2008}, this is typically done via density-density interactions with some external environment, rather than fermionic dissipation in which electrons directly tunnel into a metallic environment.

Calculating the electric conductance in the dissipation-dominated regime, we find qualitatively different behavior for IQH-SC and Laughlin FQH-SC hybrid systems. For the IQH states, the conductance saturates at low energies to a negative value if inter-edge Andreev reflection is stronger than inter-edge back-scattering, and to a positive value  
otherwise. Conversely, FQH states may display an enhancement of CAR with lowering temperatures over a wide temperature range, bounded from below by a temperature determined by the size of the system. This occurs if the effective interactions on the FQH edge are sufficiently attractive, as a result of the presence of the superconductor. Interestingly, while in the IQH case dissipation always spoils the robustness of localized zero-modes the FQH case exhibits two phases in which a CAR signal persists, one of which supports localized PZMs, see Fig.~\ref{subfig:phase_dia}. 
The distinction between the integer and fractional cases results from the suppression of fractional quasiparticle tunneling into the vortex core states \cite{shytov_tunneling_1998,chang_observation_1996,grayson_continuum_1998,grayson_resonant_2001}. Our results are congruent with the aforementioned reporting of larger CAR for fractional vs.~integer states in Ref.~\cite{Ronen_2020}.

\noindent{\it Model}---
We model the system as described in Fig.~\ref{subfig:model_illus}. A superconductor of length $L$ is embedded within the bulk of a Laughlin FQH state at filling factor $\nu = 1/m$ for odd integer $m$. We describe the QH edge in terms of two modes, converging at the end of the finger: a right-moving mode, which emanates towards the finger from a reservoir at bias voltage $V$, and a left-moving mode, which is collected by a grounded reservoir. We denote the current collected at this reservoir as $I$. 

Within the superconductor, we model the presence of vortices as a continuum of metallic states, justified by the dense spacing of the vortex core Caroli-De Gennes-Matricon states \cite{caroli_bound_1964}. We thus treat these states as metallic quasi one dimensional leads, with an effective ``Fermi velocity" corresponding to the normal NbN density of states. We assume that vortices are sufficiently prevalent such that these states are available throughout the edge. Three processes are enabled within the finger: back-scattering between the edges; crossed Andreev reflection (CAR) via the superconductor; and tunneling into the metallic states. These processes are characterized phenomenologically by tunneling amplitudes of $\Gamma$, $\Delta$, and $w$, respectively.

We define right- and left-moving boson fields, with the electron's annihilation operators being $\psi_{R/L} = e^{\pm im\phi_{R/L}}/\sqrt{2 \pi a}$, where $a$ is a short-distance cutoff of the order of the magnetic length \cite{giamarchi_quantum_2003}. The boson operator fields satisfy the commutation relations $[\phi_{R/L}(x),\phi_{R/L}(y)] = \pm i\pi \,\sgn(x-y)/m$ and $[\phi_{L}(x),\phi_{R}(y)] =i \pi /m$. The corresponding electric charge and current densities are $\hat{\rho}_{R/L}=\partial_{x}\phi_{R/L}/(2\pi)$ and $\hat{j}_{R/L}=-\partial_{t}\phi_{R/L}/(2\pi)$, respectively.

The Hamiltonian of the system is given in terms of the bosonized fields by $\mathcal{H} = H_0 + H_{\zeta} + H_{\Gamma} + H_{\Delta} +H_{w}$ \cite{wen_quantum_2004,giamarchi_quantum_2003}, where
\begin{equation}
\begin{aligned}
    \label{eq:Hamiltonian}
    H_0 = & \frac{ m}{4 \pi} \int_{-\infty}^{L} dx \bigg\{ v \left[\left( \partial_x \phi_R \right)^2 + \left( \partial_x \phi_L \right)^2\right] \\
    &\hphantom{=\frac{m}{4\pi}\int dx\,d\tau\, }
    + 2U \left( \partial_x \phi_R \right)\left( \partial_x \phi_L \right) \bigg\}, \\
    H_{\zeta} = & \sum_{\gamma = R, L}\int dx \,dy\, \zeta_\gamma ^\dagger(x,y) \left(-i v_b \partial_y \right)\zeta_\gamma (x,y),  \\
    H_{\Gamma} = & \frac{\Gamma}{\pi a^2} \int_{0}^{L} dx \cos{ \left( m \phi_R +m \phi_L \right) }, \\
    H_{\Delta} = & \frac{\Delta}{\pi a^2} \int_{0}^{L} dx \cos{ \left( m \phi_R -m \phi_L \right) }, \\
    H_{w} = & \sum_{\gamma = R, L} \int_{0}^{L} dx \left[ \frac{w}{\sqrt{a}}\, \psi_\gamma^\dagger(x)\zeta_\gamma(x,y=0) + \text{h.c.} \right].
\end{aligned}
\end{equation}
Here, $v$ and $v_b$ are Fermi velocities of the FQH edge and the metallic states, respectively, $U$ is interaction strength between the two edges, $\zeta_{R/L}$ annihilates an electron in a metallic mode that couples to the  right/left-moving edge, and we chose the units such that $\hbar=1$. These define the Luttinger velocity $u=\sqrt{v^2-U^2}$ and the Luttinger parameter $K = \sqrt{\frac{v-U}{v+U}}$. Notice that $\Gamma$, $\Delta$, $w$ and $U$ all carry units of velocity. We assume that the edges are spin polarized and the superconductor has sufficient spin-orbit coupling to induce pairing between them.

We proceed by integrating out the bath degrees of freedom. A calculation of the perturbative RG flow equations with respect to the Luttinger liquid fixed-point, $H_0$, is straightforward and yields (see App. B, \cite{SupplementaryMaterial})
\begin{equation}
    \begin{gathered}
        \begin{aligned}
        \frac{d\Delta}{d\ell} &= (2-m/K)\Delta,\qquad &&
        \frac{d\Gamma}{d\ell} &= (2-mK)\Gamma, \\
        \frac{d\alpha}{d\ell} &= (2-m\bar{K})\alpha, &&
        \frac{du}{d\ell} &= -2m\bar{K}\alpha,
        \end{aligned}
    \\
    \frac{dK}{d\ell} = m\left(\frac{\Delta}{u}\right)^{2}
                     - m K^2\left(\frac{\Gamma}{u}\right)^{2}
                     +  m\frac{\alpha}{u}(1-K^{2}),
\end{gathered}
\label{eq:RG eqns}
\end{equation}
where $\alpha = |w|^2/(2\pi v_b)$, and $\bar{K} = (K+K^{-1})/2$. 

We see that all three processes are relevant for IQH edges, and irrelevant for non-interacting FQH edges. In FQH edges, Andreev reflection becomes relevant for sufficiently attractive interactions, $K>K_c = m/2$; importantly, Andreev reflection becomes the least irrelevant process at weaker attractive interactions, $K>K_{\mathrm{CAR}}=\frac{\sqrt{3m^2 +4}-2}{m}$. 
The phase boundaries are obtained by examining the equation for $K$ and comparing the scaling of $\Delta^2$, $\Gamma^2$ and $\alpha$ which appear in this equation. In the FQH case this admits a regime, $K_{\mathrm{CAR}}<K<K_c$, that does not support localized PZMs, but can exhibit CAR at non-zero temperatures. A schematic phase diagram is shown in Fig.~\ref{subfig:phase_dia}.

\noindent{\it Temperature Dependent Perturbative Solution}---
We calculate transport properties of the system perturbatively in all three couplings. The RG equation for $K$ shows that, for IQH in the dissipation dominated regime, the system flows towards $K=1$. As such, in the IQH case, we set $K=1$.

For the perturbative calculation, we imagine dividing the finger 
into multiple segments of equal length $l$, such that each segment is close to local equilibrium. This enables us to define a local voltage for each segment, $V_{R/L}(x_n)$, $x_n = n\cdot l$. Furthermore, we assume that propagating quasiparticles lose coherence between segments, formally implemented by neglecting inter-segment correlations. In essence, we treat each segment as a quantum resistor, and the resulting network of resistors is treated classically. Motivated by the strong dissipation apparent in the experimental data, we assume that we are in the dissipation dominated regime. In this regime, we take the segment length-scale $l$ to be the dissipation length $l_d$, defined as the length scale at which most of the current dissipates into the vortices.

We proceed by writing continuity equations for the right- and left-moving densities, $\partial_t \hat{\rho}_{R/L} = -i\left[\hat{\rho}_{R/L},\mathcal{H} \right]$. Taking expectation values with respect to the unperturbed Hamiltonian $H_0$, and restricting ourselves to steady states $\left<\partial_t \hat{\rho}_{R/L}\right>=0$, we find a ``Kirchhoff-like" equation relating the currents at boundaries between segments, given at the small segment limit $l_d \ll L$ by
\begin{equation}
    \label{eq:Kirchhoff}
    \begin{aligned}
    \left< \partial_x \hat{j}_R(x)\right> &= -\left<\hat{J}_{\Gamma}(x)\right> - \left<\hat{J}_{\Delta}(x)\right> - \left<\hat{J}_{w,R}(x)\right>, \\
    \left<\partial_x \hat{j}_L(x)\right> &= -\left<\hat{J}_{\Gamma}(x)\right> + \left<\hat{J}_{\Delta}(x)\right> + \left<\hat{J}_{w,L}(x)\right>.
    \end{aligned}
\end{equation}
Here, the operators on the RHS denote the three processes that current can undergo in each segment: $\hat{J}_{\Gamma}$ describes the current per unit length that back-scatters from the right-moving to the left-moving edge; $\hat{J}_{\Delta}$ describes the Andreev current per unit length which flows from the edges into the superconductor; and $\hat{J}_{w,R/L}$ describes the current per unit length that dissipates from the right/left-moving edge to the metallic states.

The expectation values of all the operators are evaluated perturbatively to leading orders in $|\Delta|^2$,$|\Gamma|^2$, and $\alpha$. Within this treatment, the LHS of Eq.~\eqref{eq:Kirchhoff} is given by $\braket{\hat{j}_{R/L}(x_i)} = \sigma_{xy} V_{R/L}(x_i)$, where $\sigma_{xy}$ is the Hall conductance. The RHS is calculated in App. C, \cite{SupplementaryMaterial}. Due to the lack of coherence between segments, the expectation values depend on the voltage in a local manner.

At non-zero temperature and sufficiently low voltages, $e V\ll k_B T$, each quantum resistor is at the Ohmic limit. Explicit calculation then gives, for each segment, $\braket{\hat{J}_{\Gamma}(x)} = \frac{e^2}{u^2 a} A_{\Gamma} |\Gamma|^2 \left(V_R(x) - V_L(x) \right)$, $\braket{\hat{J}_{\Delta}(x)} = \frac{e^2}{u^2 a} A_{\Delta} |\Delta|^2 \left(V_R(x) + V_L(x) \right)$, and $\braket{\hat{J}_{w,R/L}(x)} = \frac{e^2}{u a}A_{w} \alpha V_{R/L}(x)$, where $A_c(T)$ with $c=\Gamma,\Delta,w$ are unitless coefficients that encode temperature dependence, and we suppress the $T$ dependence for brevity. The derivation and explicit forms of the $A_c (T)$ coefficients is given in App. C, \cite{SupplementaryMaterial} (see Table C.1). We plug these values into Eq.~\eqref{eq:Kirchhoff}, and solve it with the boundary conditions $V_R(0)=V$ (representing the incoming bias voltage) and $V_R(L)=V_L(L)$ (representing the edge of the finger, where all remaining right-movers become left-movers). The collected current is given by $I = \sigma_{xy}V_L(0)$. 

\begin{figure}[t]
    \centering
    \includegraphics[width=\columnwidth]{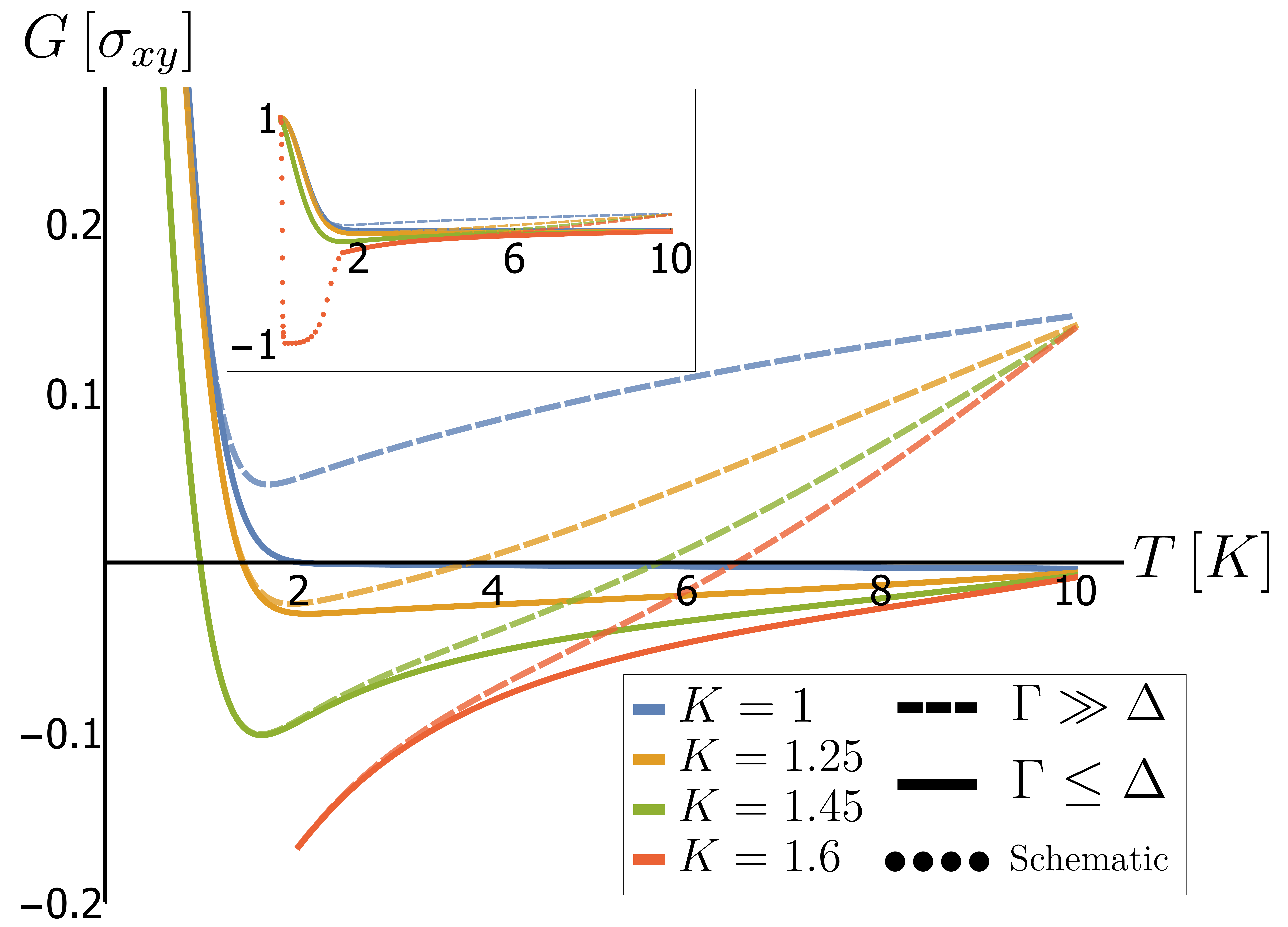}
    \caption{Differential conductance $G\equiv dI/dV$ (normalized by Hall conductance) as a function of temperature for Laughlin fractional edges ($m=3$) and various Luttinger parameters $K$, as obtained from a solution of Eq.~\eqref{eq:Kirchhoff}. The results displayed are for a system length of $L= 2 \,\upmu\mathrm{m}$, a back-scattering length of $\left( |\Gamma|^2/4\pi u^2 a\right)^{-1}= 100\,\mathrm{nm}$ (dashed lines) or  $\left( |\Gamma|^2/4\pi u^2 a\right)^{-1}=2 \,\upmu\rm{m}$ (solid lines), an Andreev-reflection length of $\left(
    |\Delta|^2/4\pi u^2 a\right)^{-1}= 1.6 \,\upmu\rm{m}$, a dissipation length of $\left(|w|^2/u v_b a\right)^{-1}= 10\,\mathrm{nm}$, a Luttinger velocity of $u=2.5\times10^4\,\mathrm{m/s}$, and a short-distance cutoff of $a=7\,\mathrm{nm}$, such that we are in the regime where bare dissipation dominates bare back-scattering and Andreev reflection. We see that, over a range of low temperatures comparable to experimental data in Ref.~\cite{Ronen_2020}, CAR increases with decreasing temperature for sufficiently attractive interactions ($K>K_{\text{CAR}}$). Inset: At sufficiently low temperatures, the finite size of the finger causes conductance to rise back to $\sigma_{xy}$. As a function of system size, this transition temperature decays as a power law for $K<K_c=1.5$, and exponentially for $K>K_c$ (schematically drawn with the dotted line). Just above this transition temperature, the $K>K_c$ case is near the full Andreev reflection fixed-point, driving the conductance to $-\sigma_{xy}$.}
    \label{fig:ResultsByTempLaughlin}
\end{figure}

The full solution for a general $L$ is given in Eq.~(C20) \cite{SupplementaryMaterial}. At the infinite finger limit, $L \rightarrow \infty$, we obtain the result 
\begin{equation}
    \label{eq:Solution_Kirchhoff}
    \frac{V_L(0)}{V} = 
    \frac{ A_{\Gamma} |\Gamma|^2 - A_{\Delta} |\Delta|^2  }
    { A_{\Gamma} |\Gamma|^2 + A_{\Delta} |\Delta|^2 + u A_{w} \alpha + \lambda },
\end{equation}
where for convenience we define $(\lambda\frac{e}{hm})^2 \equiv \left( A_{\Gamma} |\Gamma|^2 + A_{\Delta} |\Delta|^2 + u A_{w} \alpha \right)^2 - \left( A_{\Gamma} |\Gamma|^2 - A_{\Delta} |\Delta|^2 \right)^2$. As such, even in the case where all three processes are irrelevant, the low-energy behavior is dominated by the process which is \textit{least} irrelevant, with differential conductance at zero temperature taking a well-quantized value of $G/\sigma_{xy}=\pm 1, 0$. Andreev reflection becomes the least irrelevant process at $K > K_{\mathrm{CAR}}$, whereas it becomes relevant at $K > K_c$; in particular, for $m=3$ a low-energy CAR signature is obtained without localized PZMs in the range $1.19 \lesssim K<1.5$. Notice that a crossover between negative and positive conductance can occur if $\left(A_{\Gamma} |\Gamma|^2 - A_{\Delta} |\Delta|^2\right)$ changes sign at a non-zero temperature, see Fig.~\ref{fig:ResultsByTempLaughlin}.

Microscopic details, such as the coherence length ($\sim 50\textrm{nm}$ for NbN) and penetration depth (200-300nm) of the superconductor, and the metallic density of states (which is related to the $\Delta^2/E_F$ level spacing of the Caroli-De Gennes-Matricon states \cite{caroli_bound_1964}), enter the model through the three bare tunneling amplitudes As such, variance in these parameters will not affect the RG flow of Eq.~\eqref{eq:RG eqns}, nor will they effect the qualitative phase boundaries in Fig.~\ref{subfig:phase_dia}. Such varience will, however, affect the precise measured conductance of Fig.~\ref{fig:ResultsByTempLaughlin} (compare solid to dashed lines).

Finite $L$ affects the behavior at very low temperatures. At finite $L$ and $m>1$, the temperature dependence of $l_d$ defines a temperature scale at which $l_d (T_{\text{fs}}) \sim L$. It is reasonable to surmise that at $T \ll T_{\text{fs}}$ all three processes flow to zero, such that all current flows back and forth around the entire finger unperturbed, yielding $G \to \sigma_{xy}$.

\begin{figure}[t]
    \centering
    \includegraphics[width=\columnwidth]{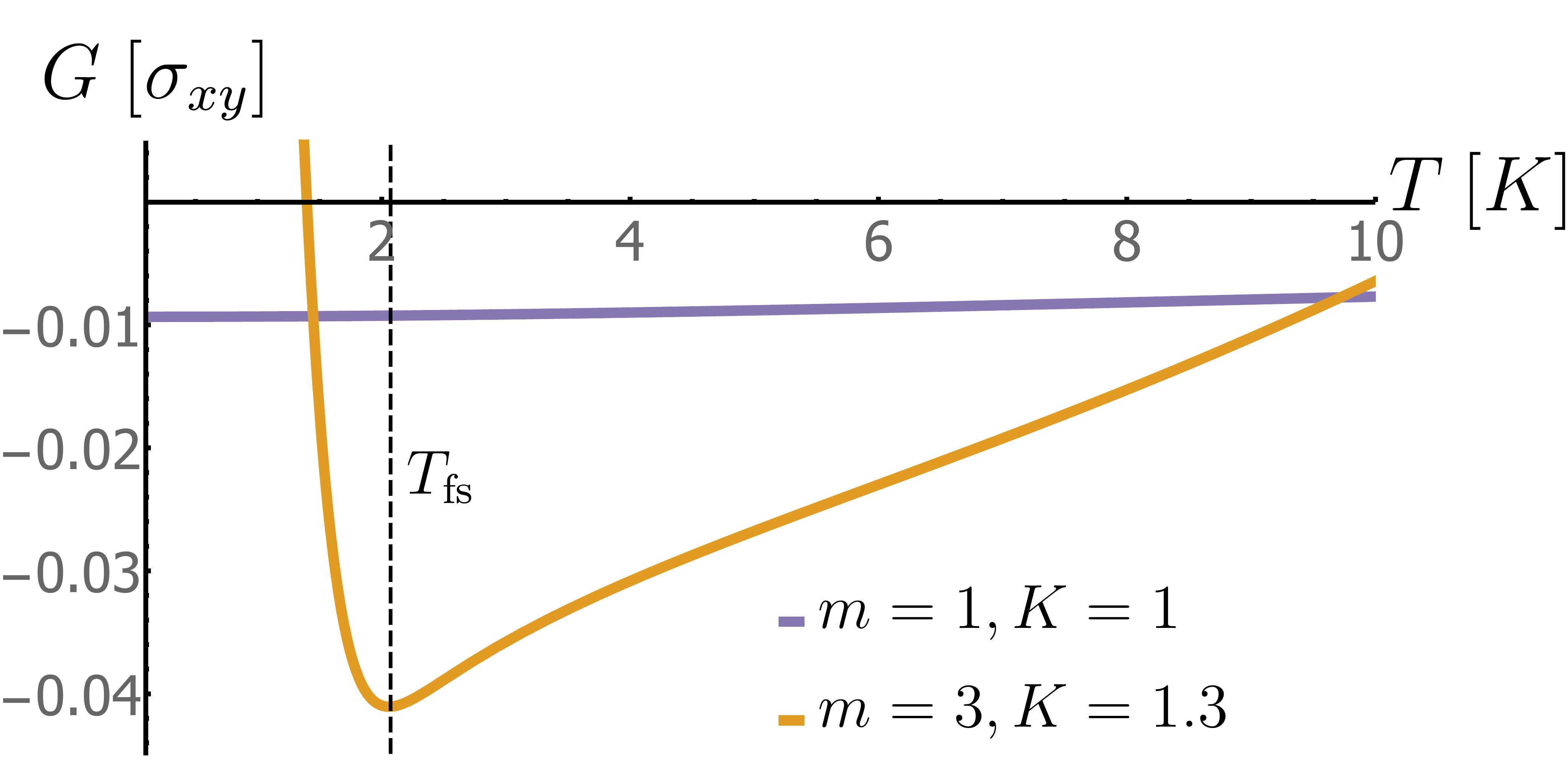}
    \caption{Differential conductance $G\equiv dI/dV$ (normalized by Hall conductance) as a function of temperature for non-interacting integer edges ($m=1$, $K=1$) (purple curve) and attractively interacting Laughlin fractional edges ($m=3, K=1.3$) (orange curve), as obtained from a solution of Eq.~\eqref{eq:Kirchhoff}. 
    We choose the same parameters as the solid lines in Fig.~\ref{fig:ResultsByTempLaughlin}, such that we are in the regime where bare dissipation dominates bare back-scattering and Andreev reflection. At the low temperature limit we show here that CAR saturates for electrons, having a very weak dependence on the temperature. For fractional quasiparticles, CAR grows monotonously with decreasing temperature over a wide range, bounded from below by $T_{\text{fs}}$ (dashed vertical line), at which point the finite size of the finger causes conductance to collapse to $\sigma_{xy}$. From the RG equations \eqref{eq:RG eqns} we see that the velocity $u$ decreases at low energies, an effect we neglected in  Eq.~\eqref{eq:Solution_Kirchhoff}; as such, the value of $T_{\text{fs}}$ shown here is an over-estimate.
    \label{fig:ResultsByTempElectrons}}
\end{figure}

In Figs.~\ref{fig:ResultsByTempLaughlin} and ~\ref{fig:ResultsByTempElectrons}, we plot the temperature dependence of the differential conductance, $G\equiv dI/dV$ for non-interacting IQH ($m=1$, $K=1$) and for $\nu=1/3$ FQH edges, respectively. We choose parameters that enable significant CAR, and give a high energy cutoff of $\sim10\,\mathrm{K}$, consistent, for example, with a bulk gap of $20\,\mathrm{K}$ found experimentally \cite{bolotin_observation_2009}. Indeed, for non-interacting IQH edges, CAR plateaus at low temperatures. For $\nu = 1/3$ edges, conversely, CAR increases as temperature decreases when interactions are sufficiently attractive, $K>K_{\mathrm{CAR}}$. This process continues until the finite size of the finger demands that all irrelevant processes halt and all current is reflected from the end of the finger.

For even stronger interactions ($K>K_c$), the Andreev pairing is relevant, and the system is driven towards a local PZM-harboring phase. At this point, Eq.~\eqref{eq:Solution_Kirchhoff} is no longer valid. For comparison, the red dotted curve in the inset of Fig.~\ref{fig:ResultsByTempLaughlin} shows a schematic temperature dependence of a power-law decay to the Andreev reflection fixed-point~\cite{schiller_predicted_2020}, with a finite size effect at low enough temperatures.

Overall, our model reproduces the temperature-dependent features seen in \cite{Ronen_2020} (in particular Fig.~5f), without requiring the proximity to the superconductor to drive the system  to a local PZM-harboring phase.

\noindent{\it Exact Solution for the $\nu=1$ case\label{sec:IQH}}---
We also ascertain the validity of our model by analyzing the case of IQH, in which our model can be written as a non-interacting electron system. The Hamiltonian in Eq.~\eqref{eq:Hamiltonian} is quadratic in fermionic operators and is exactly solvable. We obtain a non-perturbative expression for the scattering  matrix and the differential conductance, with the derivation given in App. A, \cite{SupplementaryMaterial}. We then continue to compare the temperature dependence of the conductance as calculated in the exact solution and the perturbative analysis discussed above. Moreover, we compare the conductance dependence on bias between the exact solution and the experimental data of \cite{Ronen_2020}. In the absence of dissipation and at the limit $L\to\infty$, we obtain the known result 
\begin{equation}\label{eq:SPT cond signs}
    \left.dI/dV\right|_{V=0} = (e^{2}/h) \sgn \{|\Gamma|-|\Delta|\}.
\end{equation}
These two possible values correspond to the two symmetry protected phases of Kitaev's wire model \cite{Kitaev_2001}. In the presence of dissipation we expect $|dI/dV|<e^{2}/h$. 

For finite finger length $L$ we find Tomasch oscillations \cite{Oreg_1995}, i.e., an interference effect due to reflection from both ends of the finger, with an associated energy period $E_{\text{osc.}}=hu/(2L)$. Note that these interference effects are not incorporated in the previous perturbative calculations of Eqs.~\eqref{eq:Solution_Kirchhoff}. Note that the experimental data in Ref.~\cite{Ronen_2020} shows an asymmetric bias dependence. To explain this feature, we use a complex valued $\Gamma=|\Gamma|e^{i\theta}$ which breaks bias symmetry for electrons. This allows us to qualitatively fit the experimental data of Ref.~\cite{Ronen_2020} using our model parameters, as shown in Fig.~\ref{fig:IQH_results}.

\begin{figure}[t!]
    \centering
    \includegraphics[width=\columnwidth]{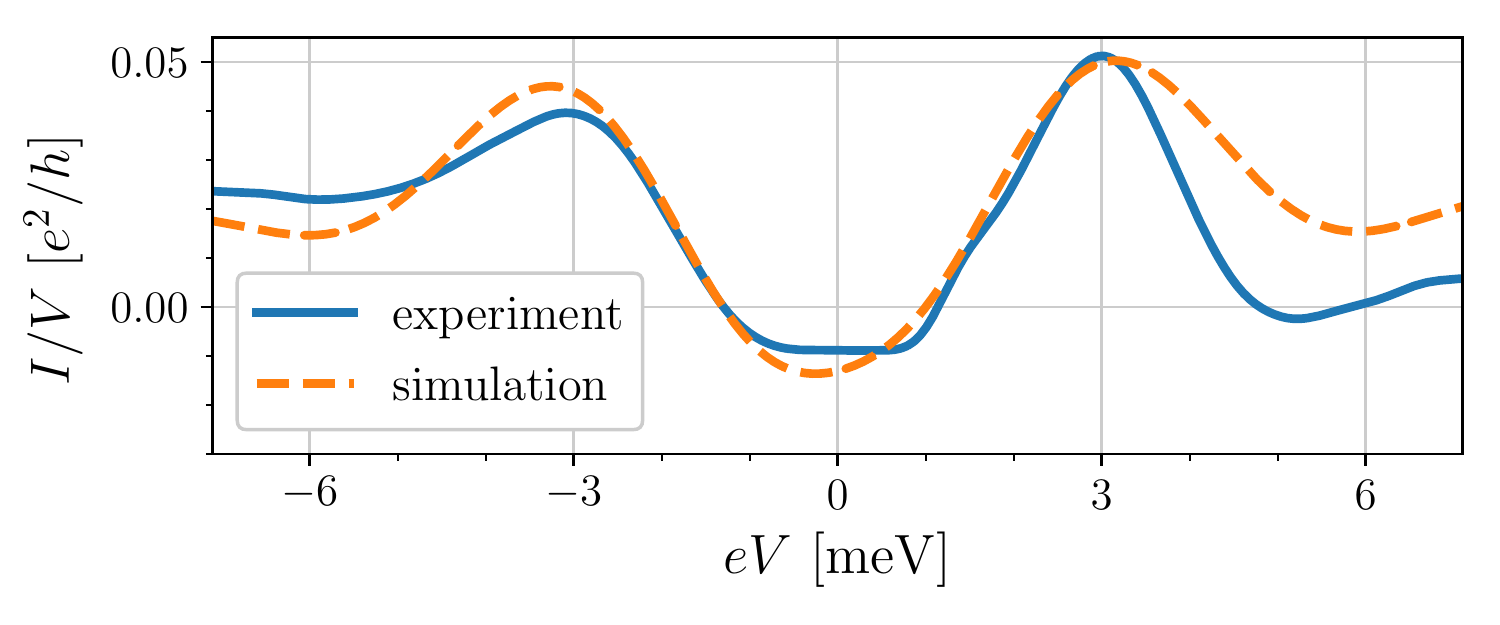}
    \caption{Comparison between exact electron conductance, $I/V$ of model \eqref{eq:Hamiltonian} in the non-interacting IQH case and the data of Ref.~\cite{Ronen_2020} (Fig.~11f). The model parameters are $|\Delta/a|=2.64\,\mathrm{meV}$, $|\Gamma/a|=2.83\,\mathrm{meV}$, $2\pi\alpha/a=2.8\,\mathrm{meV}$, $E_{\text{osc.}}=4.06\,\mathrm{meV}$ and $\theta=0.9\pi$.}
    \label{fig:IQH_results}
\end{figure}

\noindent{\it Other FQH states}---
So far, we compared the $\nu=1$ IQH state and the $\nu=1/m$ Laughlin FQH states, which feature a single edge mode. States with multiple edge modes require more delicate treatment. Qualitatively, we expect the effect of vortices on such states to depend on the relevance of electron tunneling between the particular edge and the vortex core states. Interestingly, we find that (see App. D, \cite{SupplementaryMaterial}) for $\nu=2/3$ this tunneling is marginal \cite{Kane_1994}, while for $\nu=2/5$ electron tunneling is irrelevant. This is consistent with the behavior seen in the experiment \cite{Ronen_2020} (in particular Fig.~5f), in which the conductance at $\nu=2/3$ plateaus at low temperature as in the $\nu=1$ case, while the conductance for $\nu=2/5$ is similar to that seen for $\nu=1/3$.

In summary, the ultimate cause for the qualitatively different behavior we find in various quantum Hall states is the different level of suppression of electron tunneling into metallic reservoirs~\cite{shytov_tunneling_1998}. When this suppression is strong enough, it allows for a phase that supports local PZMs even in the presence of dissipation, and allows an enhanced CAR signal relative to that obtained in the IQH case.

\noindent{\it Acknowledgments}---We thank Eduardo Fradkin, \"{O}nder G\"{u}l, Ke Huang, Philip Kim, Omri Lesser, Yuval Ronen and Jun Zhu for useful discussions,  
Evgenii Zheltonozhskii for useful comments on the original manuscript, and \"{O}nder G\"{u}l,  Philip Kim, and Yuval Ronen for graciously sharing experimental data. This work was supported by the European Union’s Horizon 2020 research and innovation programme
(grant agreements LEGOTOP No. 788715, HQMAT No. 817799, and TOPFRONT  No. 639172), the DFG 
(CRC/Transregio 183, EI 519/7-1), the BSF and NSF (2018643), the ISF
Quantum Science and Technology (2074/19). N.S.\ was supported by the Clore Scholars Programme.
\bibliographystyle{apsrev4-2}
\bibliography{main}

\newpage

\appendix
\begin{widetext}
\maketitle
\setcounter{page}{1}
\setcounter{table}{0}
\setcounter{figure}{0}
\renewcommand\thetable{\Alph{section}.\arabic{table}}
\renewcommand\thefigure{\Alph{section}.\arabic{figure}}

\section{Finger geometry in the non-interacting case}
\label{app:Smatrix}

In this appendix, we sketch the derivation of the conductance of a finger geometry in the non-interacting IQH case, leading to Fig.~\ref{fig:IQH_results} in the main text.
We do this in two steps. First, we analyze a system of counter-propagating edge states proximity coupled to a superconductor subject to a finite region with uniform coupling of superconductivity, back-scattering and dissipation, but without total reflection at the tip of the finger, as shown in Fig.~\ref{subfig:single region}. This single uniform region is suitable for determining the transport coefficients and conductance of the finger geometry in the limit of infinite finger length.
We then model a finite finger geometry by adding an additional region in which the current fully back-scatters, as shown in Fig.~\ref{subfig:two regions}. We write an analytical expression for the transfer matrix of the composite system as a composition of transfer matrices associate with the two uniform regions. We finish by finding analytical expressions for the transport coefficients and conductance of a finite finger geometry.

\floatsetup[figure]{style=plain,subcapbesideposition=top}
\begin{figure}[ht!]
    \centering
    \sidesubfloat[]{
        \includegraphics[scale=1.3]
            {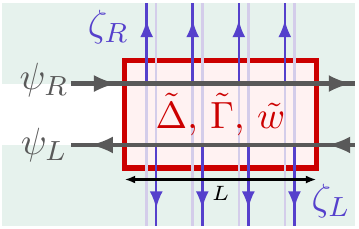}
        \label{subfig:single region}
    }
    \ 
    \sidesubfloat[]{
        \includegraphics[scale=1.3]
            {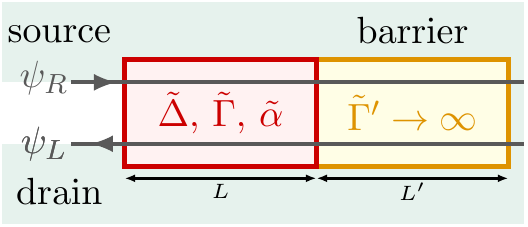}
        \label{subfig:two regions}
    }
    \caption{
    Panel~\protect\subref{subfig:single region} depicts a system of counter propagating electrons $\psi_{R/L}$ a barrier at the tip of the finger subject to superconductivity, back-scattering and electron tunneling to bath $\zeta$-electrons with uniform couplings $\tilde{\Delta}$, $\tilde{\Gamma}$ and $\tilde{w}$ respectively over a finite region $L$. Panel~\protect\subref{subfig:two regions} illustrates a system composed of two regions of uniform couplings that models a finite finger geometry. In the limit of $|\tilde{\Gamma}^\prime|\to\infty$ the right region becomes a barrier that totally reflects electrons incoming from the left. The parameter $\tilde{\alpha}\equiv|\tilde{w}|^{2}/v_{b}$ is the dissipation parameter of a system after tracing out the bath electrons.
    }
    \label{fig:two_segments}
\end{figure}

\subsection{Single uniform region}

In this subsection we consider Hamiltonian of two counter-propagating chiral electronic modes $\psi_{R/L}$ subject to superconductivity, back-scattering and electron tunneling to bath electrons $\zeta_{R/L}$ each with uniform coupling over a finite region $L$. The corresponding Hamiltonian is 
\begin{equation}
\begin{aligned}
    H^{\prime} 
        & = \intop_{-\infty}^\infty dx\, \sum_{\rho=R,L} 
            (-\rho iu\psi_{\rho}^{\dagger}\partial_{x}\psi_{\rho}) 
            + \intop_{-\infty}^\infty dy \,\intop_{|x|<L/2} dx\,\sum_{\sigma=R,L} 
            (-\sigma iv_{b}\zeta_{\sigma}^{\dagger}\partial_{y}\zeta_{\sigma}) \\
        & \hphantom{=}+\intop_{|x|<L/2}dx\,\left(
            \tilde{\Delta}\psi_{R}^{\dagger}\psi_{L}^{\dagger} 
            + \tilde{\Gamma}\psi_{R}^{\dagger}\psi_{L}+\text{h.c.}
            \right) 
            + \intop_{|x|<L/2}dx\, \sum_{\rho=R,L} \left[\tilde{w}\,
            \psi_{\rho}^{\dagger}(x)\,\zeta_{\rho}(x,y=0)+\text{h.c.}\right] 
\end{aligned}\label{eq:Hprime}
\end{equation}
where $\psi_{\rho},\zeta_{\sigma}$ satisfy fermionic anticommutation relations. As in the main text, the $\zeta$-bath electrons can be thought of as densely packed electronic wires parallel to the $y$-axis as shown in Fig.~\ref{subfig:single region}.
This system is similar to that depicted in the main text, with Hamiltonian Eq.~\eqref{eq:Hamiltonian}, in the non-interacting IQH case, however in $H^\prime$ the edge modes $\psi_\rho$ extend infinitely in both directions beyond the region $|x|<L/2$ and there is no reflective boundary condition at the end of the region that forms the tip of the finger.
If we were to ignore these differences between models, the couplings in $H^{\prime}$ would correspond to those in Hamiltonian \eqref{eq:Hamiltonian} as
\begin{equation}\label{eq:IQH coupling dict}
    \tilde{\Delta}=\Delta/a,\ \tilde{\Gamma}=\Gamma/a \quad\text{and}\quad \tilde{w}=w/\sqrt{a}.
\end{equation}

With the end goal of finding the transfer matrix and transport coefficients of $H^\prime$, we calculate the ladder operators $a_{E}^{\dagger}$ of the Hamiltonian that correspond to incoming electrons or holes from either left or right
\begin{equation}
    [H',a_{E}^{\dagger}]=Ea_{E}^{\dagger}\label{eq:raising op},
\end{equation} 
where $a_{E}^{\dagger}$ are linear combination of $\psi_{\rho}^{\dagger}(x)$, $\psi_{\rho}(x)$, $\zeta_{\sigma}^{\dagger}(x,y)$ and $\zeta_{\sigma}(x,y)$, i.e.,
\begin{equation}
\begin{aligned}
    a_{E}^{\dagger} 
        & = \intop_{x<-L/2}dx\, e^{+iEx/u}\left(
            A_{R}^{h}\psi_{R} + A_{R}^{e}\psi_{R}^{\dagger}
            \right) + e^{-iEx/u}\left(
            A_{L}^{h}\psi_{L} + A_{L}^{e}\psi_{L}^{\dagger}
            \right) \\
        & \hphantom{=} +\intop_{|x|<L/2}dx\, \sum_{\rho=R,L}\left(
            B_{\rho}^{h}(x)\psi_{\rho} + B_{\rho}^{e}(x)\psi_{\rho}^{\dagger}
            \right) 
            + \iintop_{|x|<L/2}dx\,dy\, \sum_{\sigma=R,L}\left( 
            F_{\sigma}^{h}(x,y)\zeta_{\sigma} + F_{\sigma}^{e}(x,y)\zeta_{\sigma}^{\dagger}
            \right) \\
        & \hphantom{=} +\intop_{x>+L/2}dx\, \Bigl\{ e^{+iEx/u} \left(
            C_{R}^{h}\psi_{R} + C_{R}^{e}\psi_{R}^{\dagger}
            \right) 
            + e^{-iEx/u} \left( 
            C_{L}^{h}\psi_{L} + C_{L}^{e}\psi_{L}^{\dagger}
            \right)\Bigr\}.
\end{aligned}\label{eq:ladder form}
\end{equation}
We denote the coefficients of the $\psi$-electrons in \eqref{eq:ladder form} corresponding to the area $x<-L/2$ or $x>L/2$, as two vectors
\begin{equation}
    \vec{a} = 
        \left(\begin{array}{cc|cc}
            A_{R}^{e}, & A_{R}^{h} & A_{L}^{e}, & A_{L}^{h}
        \end{array}\right)^{t}, \qquad        
    \vec{c} =\left(\begin{array}{cc|cc}
            C_{R}^{e}, & C_{R}^{h} & C_{L}^{e}, & C_{L}^{h}
        \end{array}\right)^{t},
\end{equation}
respectively. The coefficient vectors $\vec{a}$ and $\vec{c}$ encode the amplitudes of the incoming and outgoing electron or hole in the mode $a_E$ of energy $E$.
To describe an incoming $\psi$-electron or hole, and not an incoming bath electron, we demand that $a_{E}^{\dagger}$ has no support of upstream bath-electrons, i.e., $F_{\sigma}^{\chi}(\sigma y<0)=0$ with $\sigma=+$ corresponding to $\sigma=R$ (and $\sigma=-$ to $L$). 
Together with relation \eqref{eq:raising op} it is straightforward to show that the coefficient vectors $\vec{a}$ and $\vec{c}$ are related by the transfer matrix $\vec{M}$  which satisfy 
\begin{equation}
    \vec{c} =\vec{M}\vec{a},\qquad
    \vec{M} =e^{\frac{i}{u}\vec{K}E\frac{L}{2}}e^{\frac{i}{u}\vec{H}\,L}e^{\frac{i}{u}\vec{K}E\frac{L}{2}},
\end{equation}
where 
\begin{equation}
    \vec{K} \equiv 
    \left(\begin{array}{cc|cc}
        -1 & 0 & 0 & 0\\
        0 & -1 & 0 & 0\\
        \hline
        0 & 0 & 1 & 0\\
        0 & 0 & 0 & 1
        \end{array}\right), \qquad 
    \vec{H}\equiv\left(\begin{array}{cc|cc}
        \tilde{E} & 0 & -\tilde{\Gamma} & -\tilde{\Delta}\\
        0 & \tilde{E} & \tilde{\Delta}^{*} & \tilde{\Gamma}^{*}\\
        \hline
        \vphantom{\frac{\frac{s}{s}}{s}}\tilde{\Gamma}^{*} & -\tilde{\Delta} & -\tilde{E} & 0\\
        \tilde{\Delta}^{*} & -\tilde{\Gamma} & 0 & -\tilde{E}
        \end{array}\right)
\end{equation}
with $\tilde{E}=E+i|\tilde{w}|^{2}/(2v_{b})$. Notice that dissipation enters the transfer matrix only via the coefficient $\tilde{\alpha}\equiv|\tilde{w}|^{2}/v_{b}$, that is related to the $\alpha$ coefficient appearing in the main text by $\tilde{\alpha}=2\pi\alpha/a$.

The transfer matrix is given in block form by
$ \vec{M}=\left(\begin{array}{c|c}
    \vec{M}_{RR} & \vec{M}_{RL}\\
    \hline 
    \vec{M}_{LR} & \vec{M}_{LL}
\end{array}\right)$,
where
\begin{equation}\label{eq:transfer mat seg}
\begin{aligned}
\vec{M}_{RR} 
    & =e^{-\frac{i}{u}EL} 
    \begin{pmatrix}
        \frac{(c_{+}+is_{+})+(c_{-}+is_{-})}{2} 
        & -e^{i(\theta_{\Gamma}+\theta_{\Delta})} \frac{
            (c_{+}+is_{+})-(c_{-}+is_{-})}{2}
    \\
        -e^{-i(\theta_{\Gamma}+\theta_{\Delta})} \frac{
            (c_{+}+is_{+})-(c_{-}+is_{-})}{2} 
        & \frac{
            (c_{+}+is_{+})+(c_{-}+is_{-})}{2}
    \end{pmatrix},
\\
\vec{M}_{LL} 
    & =e^{\frac{i}{u}EL}
    \begin{pmatrix} 
        \frac{
            (c_{+}-is_{+})+(c_{-}-is_{-})}{2} 
        & e^{-i(\theta_{\Gamma}-\theta_{\Delta})} \frac{
            (c_{+}-is_{+})-(c_{-}-is_{-})}{2}
    \\
        e^{i(\theta_{\Gamma}-\theta_{\Delta})} \frac{
            (c_{+}-is_{+})-(c_{-}-is_{-})}{2} 
        & \frac{
            (c_{+}-is_{+})+(c_{-}-is_{-})}{2}
    \end{pmatrix},
\\
\vec{M}_{RL} 
    & =\frac{1}{2}
    \begin{pmatrix}
        -ie^{i\theta_{\Gamma}} \left(t_{+}s_{+}+t_{-}s_{-}\right) 
        & -ie^{i\theta_{\Delta}} \left(t_{+}s_{+}-t_{-}s_{-}\right)
    \\
        ie^{-i\theta_{\Delta}} \left(t_{+}s_{+}-t_{-}s_{-}\right) 
        & ie^{-i\theta_{\Gamma}} \left(t_{+}s_{+}+t_{-}s_{-}\right)
    \end{pmatrix},
\\
\vec{M}_{LR} 
    & =\frac{1}{2}
    \begin{pmatrix}
        ie^{-i\theta_{\Gamma}} \left(t_{+}s_{+}+t_{-}s_{-}\right) 
        & -ie^{i\theta_{\Delta}} \left(t_{+}s_{+}-t_{-}s_{-}\right)
    \\
        ie^{-i\theta_{\Delta}}\left(t_{+}s_{+}-t_{-}s_{-}\right) 
        & -ie^{i\theta_{\Gamma}}\left(t_{+}s_{+}+t_{-}s_{-}\right)
    \end{pmatrix},
\end{aligned}
\end{equation}
and we denote 
$t_{\pm}=(|\tilde{\Gamma}|\pm|\tilde{\Delta}|)/\tilde{E}$,
$c_{\pm}=\cos(\frac{\lambda_{\pm}L}{u})$, 
$s_{\pm}=\frac{\tilde{E}}{\lambda_{\pm}}\sin(\frac{\lambda_{\pm}L}{u})$,
$\tilde{\Delta}=|\tilde{\Delta}|e^{i\theta_{\Delta}}$ and 
$\tilde{\Gamma}=|\tilde{\Gamma}|e^{i\theta_{\Gamma}}$.
The coefficients $\pm\lambda_{+}$, $\pm\lambda_{-}$ are the eigenvalues of the matrix $\vec{H}$ and are given by  $\lambda_{\pm}=\sqrt{\tilde{E}^{2}-(|\tilde{\Gamma}|\pm|\tilde{\Delta}|)^{2}}$ with the sign of the complex square root chosen so that $\text{Im}[\lambda_{\pm}]\geq0$
and $\text{Re}[\lambda_{\pm}]$ has the same sign as $E$. 

The transport coefficients are given by
\begin{equation}\label{eq:transfer-scattering}
    \left(\begin{array}{c|c}
        \vec{T}^{+} & \vec{0}\\
        \hline 
        \vec{R}^{-} & \mathbb{I}_{2}
    \end{array}\right)
    =\left(\begin{array}{c|c}
        \mathbb{I}_{2} & \vec{R}^{+}\\
        \hline 
        \vec{0} & \vec{T}^{-}
    \end{array}\right)
    \vec{M}^{t},
\end{equation}
where $\vec{T}^{\pm}$ and $\vec{R}^{\pm}$ are the transmittance and reflectance matrices, and their entries are the transmission and reflection of electrons or holes into electrons or holes as
\begin{equation}
\vec{T}^{\pm} = 
    \begin{pmatrix}
        t_{ee}^{\pm} & t_{eh}^{\pm}\\
        t_{he}^{\pm} & t_{hh}^{\pm}
    \end{pmatrix}
\quad\text{and}\quad
\vec{R}^{\pm} = 
    \begin{pmatrix}
        r_{ee}^{\pm} & r_{eh}^{\pm}\\
        r_{he}^{\pm} & r_{hh}^{\pm}
    \end{pmatrix}.
\end{equation}
The matrices $\vec{T}^{+},\vec{R}^{+}$ correspond to the transport case of incoming electrons or holes from the left $x<-L/2$, while $\vec{T}^{-}$, $\vec{R}^{-}$ relate to the case of incoming modes from the right $x>L/2$.
The transport coefficients $\vec{T}^{\pm}$ and $\vec{R}^{\pm}$ can be written down explicitly as
\begin{align}
\vec{T}^{\pm} 
    &= \frac{e^{-\frac{i}{u}EL}}{2}
        \begin{pmatrix}
            \frac{1}{c_{+}-is_{+}}+\frac{1}{c_{-}-is_{-}} 
            & \mp e^{-i(\pm\theta_{\Gamma}+\theta_{\Delta})}        \left(
                    \frac{1}{c_{+}-is_{+}} - \frac{1}{c_{-}-is_{-}}
                \right) 
        \\ 
            \mp e^{i(\pm\theta_{\Gamma}+\theta_{\Delta})} 
                \left(
                    \frac{1}{c_{+}-is_{+}} - \frac{1}{c_{-}-is_{-}}
                \right) 
            & \frac{1}{c_{+}-is_{+}}+\frac{1}{c_{-}-is_{-}}
        \end{pmatrix},\label{eq:Tpm} \\
\vec{R}^{\pm} 
    &= \frac{e^{-\frac{i}{u}EL}}{2}
        \begin{pmatrix}
            -ie^{\mp i\theta_{\Gamma}}
                \left(
                    t_{+}\frac{s_{+}}{c_{+}-is_{+}} + t_{-}\frac{s_{-}}{c_{-}-is_{-}}
                \right) 
            & \ \mp ie^{-i\theta_{\Delta}} 
                \left(
                    t_{+}\frac{s_{+}}{c_{+}-is_{+}} - t_{-}\frac{s_{-}}{c_{-}-is_{-}}
                \right)
        \\
            \pm ie^{i\theta_{\Delta}}
                \left(
                    t_{+}\frac{s_{+}}{c_{+}-is_{+}} - t_{-}\frac{s_{-}}{c_{-}-is_{-}}
                \right) 
            & ie^{\pm i\theta_{\Gamma}} 
                \left(
                    t_{+}\frac{s_{+}}{c_{+}-is_{+}} + t_{-}\frac{s_{-}}{c_{-}-is_{-}}
                \right)
        \end{pmatrix}.\label{eq:Rpm}
\end{align}

The differential conductance of an electron at zero temperature is then given by 
\begin{equation}
    \frac{dI}{dV} = \frac{e^{2}}{h} \left(|r_{ee}^{+}|^{2}-|r_{eh}^{+}|^{2}\right) = \frac{e^{2}}{h} \left(|r_{ee}^{-}|^{2}-|r_{eh}^{-}|^{2}\right).
\end{equation}
In the limit of infinite region length, $\text{Im}[\lambda_{+}]L/u\gg1$, the transport coefficients are that of a finger of infinite length. In this limit the conductance is 
\begin{equation}
\frac{dI}{dV} 
\xrightarrow[{\text{Im}[\lambda_{+}]L/u\gg1}]{}
\frac{e^{2}}{h} \text{Re}\left[
    \frac{
        |\tilde{\Gamma}|^{2}-|\tilde{\Delta}|^{2}
    }{
        (\lambda_{+}+\tilde{E})(\lambda_{-}+\tilde{E})^{*}
    }\right].
\end{equation}
To probe the effect of small dissipation on the CAR signal we take the limits $E\ll|\tilde{\alpha}|\ll\bigl||\tilde{\Gamma}|-|\tilde{\Delta}|\bigr|$ and find
\begin{equation}
    \frac{dI}{dV}=\frac{e^2}{h}\sgn \{|\tilde{\Gamma}|-|\tilde{\Delta}|\} 
    \left\{ 1-\frac{\max(|\tilde{\Delta}|,|\tilde{\Gamma}|)}{\bigl| |\tilde{\Gamma}|+|\tilde{\Delta}|\bigr|} \cdot \frac{\tilde{\alpha}}{\bigl||\tilde{\Gamma}|-|\tilde{\Delta}|\bigr|} + O\left(\biggl(\frac{\tilde{\alpha}}{\bigl||\tilde{\Gamma}|-|\tilde{\Delta}|\bigr|}\biggr)^2\right)\right\}.
\end{equation}
The leading term in the last equation is that in Eq.~\eqref{eq:SPT cond signs}, whereas the corrections spoil the quantized conductance [as discussed in the main text below Eq.~\eqref{eq:SPT cond signs}].

\subsection{Transport coefficients of a finger geometry}

We can think of a finite finger geometry, as in Hamiltonian \eqref{eq:Hamiltonian}, as composed of two regions as shown Fig.~\ref{fig:two_segments}. The first region has couplings as per Eq.~\eqref{eq:IQH coupling dict} and length $L$ and the subsequent region is taken in the limit of infinite back-scattering $|\tilde{\Gamma}^{\prime}|\to\infty$. This second region effectively acts as a barrier and totally reflects electrons incoming from the first region. 

To compute the transport coefficient, we take the composite system with the second region bearing some finite couplings $\tilde{\Delta}^\prime$, $\tilde{\Gamma}^\prime$, $\tilde{\alpha}^\prime$ and finite length $L^\prime$. Then after computing the reflectance matrix $\vec{R}^{+}$, we will take the appropriate limits to model a barrier. 

The transfer matrix of a single region [Eq.~\eqref{eq:transfer mat seg}] allows us to write the transfer matrix of the composite system as $\vec{M}_{\text{tot}}=\vec{M}({\scriptstyle \tilde{\Gamma}^{\prime},\tilde{\Delta}^{\prime},\tilde{\alpha}^{\prime}}) \vec{M}({\scriptstyle \tilde{\Gamma},\tilde{\Delta},\tilde{\alpha}})$. We can then find the $\vec{T}^{\pm}$, $\vec{R}^{\pm}$ via Eq.~\eqref{eq:transfer-scattering}. In the limit of $|\tilde{\Gamma}^{\prime}|\to\infty$ we find that $\vec{T}^{+}=0$ and 
\begin{equation}
\begin{aligned}
\vec{R}^{+} 
    & =\frac{ie^{-\frac{2i}{u}EL}}{\mathcal{A}}\left\{ 
        \left(\begin{array}{cc}
            -e^{-i\theta_{\Gamma}}(c_{+}s_{-}t_{-}+c_{-}s_{+}t_{+}) 
            & -ie^{-i\theta_{\Delta}}(t_{+}-t_{-})s_{+}s_{-}
        \\
            -ie^{i\theta_{\Delta}}(t_{+}-t_{-})s_{+}s_{-} 
            & e^{i\theta_{\Gamma}}(c_{+}s_{-}t_{-}+c_{-}s_{+}t_{+})
        \end{array}\right)\right.
    \\
    & \hphantom{=}+\cos\theta
        \left(\begin{array}{cc}
            -e^{-i\theta_{\Gamma}}    
                \left(c_{+}c_{-}+s_{+}s_{-}(1+t_{+}t_{-})\right) 
            & ie^{-i\theta_{\Delta}}(c_{+}s_{-}-c_{-}s_{+})
        \\
            -ie^{i\theta_{\Delta}}(c_{+}s_{-}-c_{-}s_{+}) 
            & e^{i\theta_{\Gamma}}\left(
                c_{+}c_{-}+s_{+}s_{-}(1+t_{+}t_{-})\right)
        \end{array}\right)
    \\
    & \hphantom{=}+\left.\sin\theta
        \left(\begin{array}{cc}
            -ie^{-i\theta_{\Gamma}} \frac{
                c_{+}^{2} + s_{+}^{2}(1-t_{+}^{2}) + c_{-}^{2} + s_{-}^{2}(1-t_{-}^{2})
                }{2} 
            & ie^{-i\theta_{\Delta}} \frac{
                c_{+}^{2} + s_{+}^{2}(1-t_{+}^{2}) - c_{-}^{2} - s_{-}^{2}(1-t_{-}^{2})
                }{2}
        \\
            ie^{i\theta_{\Delta}} \frac{
                c_{+}^{2} + s_{+}^{2}(1-t_{+}^{2}) - c_{-}^{2} - s_{-}^{2}(1-t_{-}^{2})
                }{2} 
            & -ie^{i\theta_{\Gamma}} \frac{
                c_{+}^{2} + s_{+}^{2}(1-t_{+}^{2}) + c_{-}^{2} + s_{-}^{2}(1-t_{-}^{2})
                }{2}
        \end{array}\right)
    \right\} ,
\end{aligned}
\end{equation}
where $\theta=\theta_{\Gamma}-\theta_{\Gamma'}$ is the relative
phase between the back-scattering coefficients of the $L$ and $L^{\prime}$
segments, and 
\begin{equation}
\mathcal{A} \equiv(c_{+}-is_{+})(c_{-}-is_{-})+t_{+}t_{-}s_{+}s_{-} 
 +\cos\theta\left[c_{-}s_{+}t_{+}+c_{+}s_{-}t_{-}-i(t_{+}+t_{-})s_{+}s_{-}\right].
\end{equation}
We then find the conductance of the finger at zero temperature to be $G_0(E) = \frac{e^{2}}{h} \left(|r_{ee}^{+}|^{2}-|r_{eh}^{+}|^{2}\right)$. For finite temperature we obtain the conductance this expression with the Fermi-Dirac distribution, $I/V=\int dE\, G_0(E)\big\{f_{\mathrm{FD}}(E+eV,T) \\  -f_{\mathrm{FD}}(E,T) \big\}/V $. By fitting to the experimental data of \cite{Ronen_2020} we obtain Fig.~\ref{fig:IQH_results} of the main text.

\section{Renormalization group flow}
\label{app:RG}

In this section we elaborate on the derivation of the RG equations appearing in the main text, Eq.~\eqref{eq:RG eqns}. Our starting point is the quantum partition function $Z=\int \mathcal{D}\Phi\,e^{- S[\Phi]}$, where $S$ is the action of our model analytically continued to imaginary time ($t\to -i\tau$) and $\Phi$ are the dynamical fields, namely $\phi_{R/L}$ and $\zeta_{R/L}$. Since the model Hamiltonian in Eq.~\eqref{eq:Hamiltonian} is quadratic in the $\zeta$-fermions so is $S$, and the $\zeta$ fields can be integrated explicitly to yield the effective action $S_\text{eff}[\phi_R,\phi_L] = S_{0} + S_{\Delta} + S_{\Gamma} + S_{\alpha}$, where
\begin{equation}
\begin{aligned}
    S_{0} &= \frac{m}{4\pi}\int dx\,d\tau\,\left[\partial_{x}\phi_{R}(i\partial_{\tau}+v\partial_{x})\phi_{R}+\partial_{x}\phi_{L}(-i\partial_{\tau}+v\partial_{x})\phi_{L}+2U\partial_{x}\phi_{R}\partial_{x}\phi_{L}\right], \\
    S_{\Delta} & =\int dx\,d\tau\,\frac{\Delta}{\pi a^{2}}\cos(m\phi_{R}-m\phi_{L}), \\
    S_{\Gamma} &= \int dx\,d\tau\,\frac{\Gamma}{\pi a^{2}}\cos(m\phi_{R}+m\phi_{L}), \\
    S_{\alpha} &= \int dx\,d\tau_{1}\,d\tau_{2}\,\frac{-\alpha}{2\pi a^{2}}\left(\frac{e^{-im\phi_{R}(\tau_{1},x)}e^{im\phi_{R}(\tau_{2},x)}}{\tau_{1}-\tau_{2}}+\frac{e^{im\phi_{L}(\tau_{1},x)}e^{-im\phi_{L}(\tau_{2},x)}}{\tau_{1}-\tau_{2}}\right).
\end{aligned}
\end{equation}
Note that the effective action includes a term that is non-local in time. We derive the RG equations by treating $S_{0}$ as the fixed-point perturbed by $S_{\Delta} + S_{\Gamma} + S_{\alpha}$. We follow the minimal subtraction scheme with a short distance cutoff $a$ (see for example Ref.~\onlinecite{Cardy_1996}). The RG equations are completely determined by the conformal field theory structure of the fixed-point $S_{0}$, and in particular by the scaling dimensions of the perturbing fields and their operator product expansions.
The perturbing fields in $S_\text{eff}$ are
\begin{equation}
\begin{aligned}
    \cos(m\phi_{R}(\tau,x)-m\phi_{L}(\tau,x)),\quad
    \cos(m\phi_{R}(\tau,x)+m\phi_{L}(\tau,x))\quad \text{and}\quad
    e^{\mp im\phi_{R/L}(\tau_{1},x)}e^{\pm im\phi_{L/R}(\tau_{2},x)},
\end{aligned}
\end{equation}
and they have the scaling dimensions $m/K$, $m K$ and $m(K+K^{-1})/2$ respectively (as in the main text $K=\sqrt{\frac{v-U}{v+U}}$ and $u=\sqrt{v^2-U^2}$). From this we deduce the RG equations for $\Delta$, $\Gamma$ and $\alpha$ as in Eqs.~\eqref{eq:RG eqns}. Note that for the $\alpha$-equation we need to include the scaling dimension of the coefficient $1/(\tau_1-\tau_2)$ to obtain the correct flow equation.

We also obtain a correction coming from the cutoff shift $a\to a e^{\delta \ell}$ appearing in the $S_{\alpha}$,
\begin{equation}
    \delta S_{\text{corr.}} = \intop_{a<u|\tau_{1}-\tau_{2}|<a e^{\delta \ell}}dx\,d\tau_{1}\,d\tau_{2}\,\frac{-\alpha}{2\pi a^{2}}\left(\frac{e^{-im\phi_{R}(\tau_{1},x)}e^{im\phi_{R}(\tau_{2},x)}}{\tau_{1}-\tau_{2}}+\frac{e^{im\phi_{L}(\tau_{1},x)}e^{-im\phi_{L}(\tau_{2},x)}}{\tau_{1}-\tau_{2}}\right).
\end{equation}
This can be simplified using the operator product expansion
\begin{equation}\label{eq:OPE}
    e^{\mp i m\phi_{R/L}(\tau_{1},x_1)}e^{\pm i m\phi_{L/R}(\tau_{2},x_2)} = \frac{a^{m\bar{K}}}{z_{\pm}^{m(\bar{K}+1)/2}z_{\mp}^{m(\bar{K}-1)/2}}\left(1+\mathcal{F}_{\pm}+\frac{1}{2}\normalorder{\mathcal{F}^{2}_{\pm}}+\cdots\right)
\end{equation}
where $z_{\pm}=u(\tau_{1}-\tau_{2})\mp i(x_{1}-x_{2})$, the $(\cdots)$ are less relevant terms,  $\normalorder{\cdot}$ is the normal ordering with respect to the ground state of the ground state of $S_{0}$, and 
\begin{equation}
    \mathcal{F}_{\pm} = \frac{m}{2}
        \begin{pmatrix}
            z_{\pm} \\
            z_{\mp}
        \end{pmatrix}^t
        \begin{pmatrix}
            1+\bar{K} & \frac{K^{-1}-K}{2}\\
            1-\bar{K} & \frac{K-K^{-1}}{2}
        \end{pmatrix}
        \begin{pmatrix}
            \partial_{x}\phi_{R/L}(\frac{\tau_{1}+\tau_{2}}{2},\frac{x_{1}+x_{2}}{2})\\
            \partial_{x}\phi_{L/R}(\frac{\tau_{1}+\tau_{2}}{2},\frac{x_{1}+x_{2}}{2})
        \end{pmatrix}.
\end{equation}
Inserting the operator product in the correction terms we find that 
\begin{equation}
    \delta S_{\text{corr.}} = \delta\ell\int dx\,d\tau\, \frac{m}{4\pi}\normalorder{
        \begin{pmatrix}
            \partial_{x}\phi_{R}\\
            \partial_{x}\phi_{L}
        \end{pmatrix}^t
        \begin{pmatrix}
            -m\alpha(K^2+K^{-2}) & m\alpha(K^2-K^{-2})\\
            m\alpha(K^{2}-K^{-2}) & -m\alpha(K^2+K^{-2})
        \end{pmatrix}
        \begin{pmatrix}
            \partial_{x}\phi_{R}\\
            \partial_{x}\phi_{L}
        \end{pmatrix}}
        +\cdots
\end{equation}
where $\cdots$ includes a constant term and contributions from the terms omitted in Eq.~\eqref{eq:OPE}. From the $2\times 2$ matrix in the last expression,  $dv/d\ell$ and $dV/d\ell$ are readily read as the diagonal and off-diagonal terms respectively. The corresponding flow equations $dK/d\ell$ and $du/d\ell$ are as in Eq.~\eqref{eq:RG eqns}.

\section{Perturbative calculation of correlations}
\label{app:Perturbative}
In this appendix, we elaborate on the calculations leading to the perturbative calculation of the conductance leading to Eq.~\eqref{eq:Solution_Kirchhoff} in the main text. We work in natural units $\hbar = k_{\text{B}} = 1$.

\subsection{Kirchhoff-like equations}
\label{subsec:Continuity}
Working in the Heisenberg picture, we begin by writing a continuity equation by commuting the density operators $\hat{\rho}_{R/L}(x,t)=\partial_x \phi_{R/L}(x,t)/(2\pi)$ with the Hamiltonian $\mathcal{H}$ as given in Eq.~\eqref{eq:Hamiltonian}. We use the bosonic commutation relations to obtain the following commutators for $0<x<L$,

\begin{align}
    \label{eq:Continuity_equation}
    e \partial_t \hat{\rho}_{R/L}(x,t) = & -ie \left[ \hat{\rho}_{R/L}(x,t),\mathcal{H} \right] \\
    = & \mp \partial_x \hat{j}_{R/L}(x,t) \mp e \frac{\Gamma}{\pi a^2} \sin {\left(m(\phi_R(x,t)+\phi_L(x,t)) \right)} 
    - e \frac{\Delta}{\pi  a^2} \sin{ \left(m(\phi_R(x,t)-\phi_L(x,t)) \right)} \nonumber \\
    - & i e \left( w \psi_{R/L}^\dagger(x,t) \zeta_{R/L}(x,y=0,t)- w^*\zeta^\dagger_{R/L}(x,y=0,t) \psi_{R/L}(x,t) \right). \nonumber
\end{align}

We identify $\hat{j}_{R/L}(x,t) = \mp i e \left[ \hat{\rho}_{R/L}(x,t),H_0 \right]$, where we define the direction in which current flows for each edge as positive. We define the back-scattering, Andreev, and dissipating currents \textbf{per unit length} as
\begin{subequations}
\begin{align}
    \label{eq:Define_Densities}
    \hat{J}_{\Gamma}(x,t) = & e \frac{\Gamma}{\pi a^2} \sin {\left(m(\phi_R(x,t)+\phi_L(x,t)) \right)}, \\
    \hat{J}_{\Delta}(x,t) = & \; e \frac{\Delta}{\pi  a^2} \sin{ \left(m(\phi_R(x,t)-\phi_L(x,t)) \right)}, \\ 
    \hat{J}_{w,R/L}(x,t) = & \; i e \left( w \psi_{R/L}^\dagger(x,t) \zeta_{R/L}(x,y=0,t)- w^*\zeta^\dagger_{R/L}(x,y=0,t) \psi_{R/L}(x,t) \right),
\end{align}
\end{subequations}
respectively. Taking expectation values on all operators with respect to the unperturbed Hamiltonian $H_0$, we focus on steady state solutions, taking $\partial_t \hat{\rho}_{R/L} (x,t) = 0$, and study the operators on the right-hand-side of Eq.~\eqref{eq:Continuity_equation}. 
This gives us two point-like, ``Kirchhoff-like" equations imposing local current conservation, in the form of Eq.~\eqref{eq:Kirchhoff} of the main text (repeated here),
\begin{equation}
    \begin{aligned}
    \left< \partial_x \hat{j}_R(x)\right> &= -\left<\hat{J}_{\Gamma}(x)\right> - \left<\hat{J}_{\Delta}(x)\right> - \left<\hat{J}_{w,R}(x)\right>, \\
    -\left<\partial_x \hat{j}_L(x)\right> &= +\left<\hat{J}_{\Gamma}(x)\right> - \left<\hat{J}_{\Delta}(x)\right> - \left<\hat{J}_{w,L}(x)\right>.
    \end{aligned}
\end{equation}

We now integrate these equations over a single segment, $x_i<x<x_{i+1}$. This leads to a segmented version of Eq.~\eqref{eq:Kirchhoff},
\begin{equation}
    \begin{aligned}
    \left<\hat{j}_{R}(x_{i+1})\right> - \left<\hat{j}_{R}(x_{i})\right> &= -\left<\hat{J}_{\Gamma}^{(i)}\right> - \left<\hat{J}_{\Delta}^{(i)}\right> - \left<\hat{J}_{w,R}^{(i)}\right>, \\
    \left<\hat{j}_{L}(x_{i})\right> - \left<\hat{j}_{L}(x_{i+1})\right> &= 
    +\left<\hat{J}_{\Gamma}^{(i)}\right> -
    \left<\hat{J}_{\Delta}^{(i)}\right> - \left<\hat{J}_{w,L}^{(i)}\right>,
    \end{aligned}
    \label{eq:Kirchhoff_segmented}
\end{equation}
where we define the operator for the current which is back-scattered, is Andreev reflected, or is dissipated in the $i$-th segment  as  $\hat{J}_{c}^{(i)}\equiv\int_{x_i}^{x_{i+1}}dx\, \hat{J}_{c}(x)$, where $c$ takes the values  $\Gamma,\Delta$, and $w,R/L$, respectively. We note that at the limit of infinitesimal segment length, this is equivalent to Eq.~\eqref{eq:Kirchhoff}.

\subsection{Calculating expectation values}
\label{subsec:Expectation}

We now take expectation values of these operators. To leading order in perturbation theory, these are given by
\begin{subequations}
\begin{align}
    \label{eq:ExpectationValues_Kubo}
    \left< \hat{J}_{\Gamma}^{(i)}(t) \right> = 
    & -i \int_{x_i}^{x_{i+1}}dx \int_0^L dx^\prime \int_{-\infty}^{t}dt^\prime 
    \left< \left[ \hat{J}_{\Gamma}(x,t), H_{\Gamma}(x^\prime,t^\prime)\right] \right>, \\
    \left< \hat{J}_{\Delta}^{(i)}(t) \right> = 
    & -i \int_{x_i}^{x_{i+1}}dx \int_0^L dx^\prime \int_{-\infty}^{t}dt^\prime 
    \left< \left[ \hat{J}_{\Delta}(x,t), H_{\Delta}(x^\prime,t^\prime)\right] \right>, \\\left< \hat{J}_{w,R/L}^{(i)}(t) \right> = 
    & -i \int_{x_i}^{x_{i+1}}dx \int_0^L dx^\prime \int_{-\infty}^{t}dt^\prime 
    \left< \left[ \hat{J}_{w,R/L}(x,t), H_{w}(x^\prime,t^\prime)\right] \right>.
\end{align}
\end{subequations}
Here, all expectation values are taken with respect to the unperturbed Hamiltonian, $H_0$ (see Eq.~\eqref{eq:Hamiltonian}), the operators in the right hand side are written in the interaction picture with respect to $H_0$, and we drop terms that result in a vanishing expectation value, such as $\left< \left[ \hat{J}_{\Gamma}(x,t), H_{\Delta}(x^\prime,t^\prime)\right] \right>$. Using the explicit forms of the operators in Eq.~\eqref{eq:Define_Densities}, we keep only charge conserving terms, and utilize the fact that we're only calculating DC-quantities to extend the integration domain, obtaining
\begin{subequations}
\label{eq:ExpectationValues_Explicit}
    \begin{multline}
    \left< \hat{J}_{\Gamma}^{(i)}(t) \right> =
    -e \frac{\Gamma^2}{4\left(\pi a^2 \right)^2}
    \int_{x_i}^{x_{i+1}}dx \int_0^L dx^\prime \int_{-\infty}^{\infty}dt^\prime 
    \left< e^{im \left( \phi_R(x,t)+\phi_L(x,t)\right)} 
    e^{-im \left( \phi_R(x^\prime,t^\prime)+\phi_L(x^\prime,t^\prime)\right)} \right.\\
    \left.   -  e^{-im \left( \phi_R(x,t)+\phi_L(x,t)\right)}
    e^{im \left( \phi_R(x^\prime,t^\prime)+\phi_L(x^\prime,t^\prime)\right)}\right>, 
    \end{multline}
    \begin{multline}
    \left< \hat{J}_{\Delta}^{(i)}(t) \right> = 
     -e \frac{\Delta^2}{4\left(\pi a^2 \right)^2}
    \int_{x_i}^{x_{i+1}}dx \int_0^L dx^\prime \int_{-\infty}^{\infty}dt^\prime 
    \left< e^{im \left( \phi_R(x,t)-\phi_L(x,t)\right)} 
    e^{-im \left( \phi_R(x^\prime,t^\prime)-\phi_L(x^\prime,t^\prime)\right)} \right.\\
    \left.  -   e^{-im \left( \phi_R(x,t)-\phi_L(x,t)\right)}
    e^{im \left( \phi_R(x^\prime,t^\prime)-\phi_L(x^\prime,t^\prime)\right)}\right>, 
    \end{multline}
    \begin{multline}
    \left< \hat{J}_{w,R/L}^{(i)}(t) \right> = 
    e |w|^2
    \int_{x_i}^{x_{i+1}}dx \int_0^L dx^\prime \int_{-\infty}^{\infty}dt^\prime 
    \left< \psi^\dagger_{R/L}(x,t)\zeta_{R/L}(x,y=0,t)\zeta^\dagger_{R/L}(x^\prime,y=0,t^\prime)\psi_{R/L}(x^\prime,t^\prime) \right. \\
    - \left. \zeta^\dagger_{R/L}(x,y=0,t)\psi_{R/L}(x,t)\psi^\dagger_{R/L}(x^\prime,t^\prime)\zeta_{R/L}(x^\prime,y=0,t^\prime) \right>.
    \end{multline}
\end{subequations}

We remind the reader that $\psi_{R/L} = e^{\pm im\phi_{R/L}}/\sqrt{2 \pi a}$. Up to this point, we've made no assumptions on the system, and obtained equations that are correct generically. Now to enable calculation of these expectation values, we employ our assumption that the finger is split into segments of length $l$. Within each segment, each edge equilibrates to a voltage of $V_{R/L}(x_i) \equiv V_{R/L}(x_i\leq x \leq x_{i+1})$, where $x_i = i*l$. Furthermore, coherent processes occur only within each segment, and not between segments. 

The mechanism that determines $l$ depends on the many length scales within the system that can lead to decoherence. These include: the dissipation length $l_d$, defined as the length scale at which most of the current dissipates into the vortices; the coherence length $l_\phi$; the back-scattering/Andreev lengths, defined as the length scale at which most of the current is back-scattered/Andreev reflected between edges in the absence of dissipation; and a characteristic length scale of disorder in the system. We assume that dissipation dominates all these processes, and hence will eventually take $l=l_d$; however, we keep the calculations general for the time being.

This assumption affects all expectation values necessary for the calculation of Eq.~\eqref{eq:Kirchhoff_segmented}. When taking expectation values of the edge-current operators on the left-hand-side, we trivially obtain $\braket{\hat{j}_{R/L}(x_i)} = \sigma_{xy} V_{R/L}(x_i)$, where $\sigma_{xy}= e^2/h m$ is the Hall conductance. Furthermore, in the expectation values of the inter-edge processes, given in Eq.~\eqref{eq:ExpectationValues_Explicit}, the effects are two-fold. First, the limits of integration change from $\int_{0}^{L}dx^\prime$ to $\int_{x_i}^{x_{i+1}}dx^\prime$. Second, the expectation values in the integrands are obtained by moving to a chiral basis which diagonalizes the interacting Hamiltonian $H_0$, and incorporating the bias voltages $V_{R/L}(x_i)$ through the appropriate gauge transformations. Expectation values are then given by standard correlation functions for Luttinger liquids \cite{giamarchi_quantum_2003}, resulting in
\begin{subequations}
    \begin{align}
    \label{eq:ExpectationValues_AfterCorrelations}
    \left< \hat{J}_{\Gamma}^{(i)}(t) \right> = 
      \frac{-i e \Gamma^2}{2\left(\pi a^2 \right)^2} &
    \int\displaylimits_{x_i}^{x_{i+1}}dx dx^\prime \int\displaylimits_{-\infty}^{\infty}dt^\prime 
    \frac{\sin{\left( K \left(e V_R(x_i) + e V_L(x_i) \right) \frac{x-x^\prime}{u} - \left(e V_R(x_i) - e V_L(x_i) \right) (t - t^\prime - i\varepsilon)\right)}}
    {\left[ \left( \frac{u}{\pi T a}\right)^2 
    i \sinh{\left(\pi T \frac{-(x-x^\prime)+u(t-t^\prime - i\varepsilon)}{u}\right)} 
    i \sinh{\left(\pi T \frac{(x-x^\prime)+u(t-t^\prime - i\varepsilon)}{u}\right)}\right]^{mK}} \\
    \left< \hat{J}_{\Delta}^{(i)}(t) \right> = 
      \frac{i e \Delta^2}{2\left(\pi a^2 \right)^2} &
    \int\displaylimits_{x_i}^{x_{i+1}}dx dx^\prime \int\displaylimits_{-\infty}^{\infty}dt^\prime 
    \frac{\sin{\left( \frac{1}{K}\left(e V_R(x_i) - e V_L(x_i) \right) \frac{x-x^\prime}{u} + \left(e V_R(x_i) + e V_L(x_i) \right) (t - t^\prime - i\varepsilon)\right)}}
    {\left[ \left( \frac{u}{\pi T a}\right)^2 
    i \sinh{\left(\pi T \frac{-(x-x^\prime)+u(t-t^\prime - i\varepsilon)}{u}\right)} 
    i \sinh{\left(\pi T \frac{(x-x^\prime)+u(t-t^\prime - i\varepsilon)}{u}\right)}\right]^{m/K}} \\
    \left< \hat{J}_{w,R/L}^{(i)}(t) \right> & = 
     \frac{i e |w|^2}{2 \pi^2 a^3} \frac{u}{v_b}
    \int\displaylimits_{x_i}^{x_{i+1}}dx dx^\prime \int\displaylimits_{-\infty}^{\infty}dt^\prime \frac{\sin{\left( e V_{R/L}(x_i)(t-t^\prime - i\varepsilon)\right)}}{\left[ \left( \frac{u}{\pi T a}\right) 
    i \sinh{\left(\pi T \frac{u(t-t^\prime - i\varepsilon)}{u}\right)} \right]^{m \bar{K}+1}} \delta \left( x - x^\prime \right)
    \end{align}
\end{subequations}
where $\varepsilon\rightarrow 0^{+}$ is an infinitesimal time cutoff. We remind the reader that $\bar{K} = (K+K^{-1})/2$, as defined after Eq.~\eqref{eq:RG eqns}.

These equations can be simplified using some basic algebraic maneuvers. First, we move to center of mass coordinates, $x_+ \equiv x+x^\prime - 2l$, $x_- \equiv x - x^\prime, t_- \equiv t-t^\prime$. We immediately see that $x_+$ does not appear in the integrand, so it can be integrated on easily to give
\begin{equation*}
    \int\displaylimits_{x_i}^{x_{i+1}}dx dx^\prime \int\displaylimits_{-\infty}^{\infty}dt^\prime f(x_-,t_-)= \int\displaylimits_{-l}^{l}dx_- \int\displaylimits_{-\infty}^{\infty}dt_{-} 
    \left( l - |x_-| \right) f(x_-,t_-).
\end{equation*}
We now define unitless variables,
\begin{equation*}
    \tilde{x} \equiv \frac{x_-}{l}; \; \tilde{t} \equiv \frac{u t_-}{l}; \; \tilde{V}_{R/L}(x_i) \equiv \frac{eV_{R/L}(x_i)l}{u}; \; \tilde{T} \equiv \frac{T l}{u},
\end{equation*}
to obtain the expressions
\begin{subequations}
\begin{align}
    \left< \hat{J}_{\Gamma}^{(i)}(t) \right> = 
      \frac{-i e \Gamma^2}{2 \pi^2 u a } &\left( \frac{l}{a} \right)^{3-2mK} 
    \int\displaylimits_{-1}^{1} d\tilde{x} \int\displaylimits_{-\infty}^{\infty} d\tilde{t}
    \frac{\left( 1 - |\tilde{x}| \right) \sin{\left( K \left(\tilde{V}_R(x_i) + \tilde{V}_L(x_i) \right) \tilde{x} - \left(\tilde{V}_R(x_i) - \tilde{V}_L(x_i) \right) (\tilde{t} - i\varepsilon)\right)}}
    {\left[ \left( \frac{1}{\pi \tilde{T}}\right)^2 
    i \sinh{\left(\pi \tilde{T} \left( -\tilde{x}+\tilde{t}- i\varepsilon\right) \right)} 
     i \sinh{\left(\pi \tilde{T} \left( \tilde{x}+\tilde{t}- i\varepsilon\right) \right)}  \right]^{mK}} \\
    \left< \hat{J}_{\Delta}^{(i)}(t) \right> = 
      \frac{i e \Delta^2}{2 \pi^2 u a } &
      \left( \frac{l}{a} \right)^{3-2m/K} 
    \int\displaylimits_{-1}^{1} d\tilde{x} \int\displaylimits_{-\infty}^{\infty} d\tilde{t}
    \frac{\left( 1 - |\tilde{x}| \right) \sin{\left( \frac{1}{K}\left(\tilde{V}_R(x_i) - \tilde{V}_L(x_i) \right) \tilde{x} + \left(\tilde{V}_R(x_i) + \tilde{V}_L(x_i) \right) (\tilde{t} - i\varepsilon)\right)}}
    {\left[ \left( \frac{1}{\pi \tilde{T}}\right)^2 
    i \sinh{\left(\pi \tilde{T} \left( -\tilde{x}+\tilde{t}- i\varepsilon\right) \right)} 
     i \sinh{\left(\pi \tilde{T} \left( \tilde{x}+\tilde{t}- i\varepsilon\right) \right)}  \right]^{m/K}} \\
    \left< \hat{J}_{w,R/L}^{(i)}(t) \right> & = 
     \frac{i e |w|^2}{2 \pi^2 v_b a} \left( \frac{l}{a} \right)^{1-m\bar{K}} 
     \int\displaylimits_{-\infty}^{\infty}d\tilde{t} \frac{\sin{\left( \tilde{V}_{R/L}(x_i)(\tilde{t} - i\varepsilon)\right)}}{\left[ \left( \frac{1}{\pi \tilde{T}}\right)
    i \sinh{\left(\pi \tilde{T} \left( \tilde{t}- i\varepsilon\right) \right)} 
     \right]^{m \bar{K}+1}}.
\end{align}
\end{subequations}
We treat the infinitesimal cutoffs, as well as potential divergences, by a change of variables in the complex plane, defining $\tilde{y}\equiv \pi \tilde{T}\left(\tilde{t}-i\varepsilon+i/2\tilde{T} \right)$ \cite{martin_noise_2005}. The three integrals are now given by
\begin{subequations}
\begin{align}
    \left< \hat{J}_{\Gamma}^{(i)}(t) \right> = 
      \frac{-i e \Gamma^2}{2 \pi^2 u a } &\left( \frac{l}{a} \right)^{3-2mK} 
    \int\displaylimits_{-1}^{1} d\tilde{x} \int\displaylimits_{-\infty-i\varepsilon+\frac{i \pi}{2}}^{\infty-i\varepsilon+\frac{i \pi}{2}} \frac{d \tilde{y}}{\pi \tilde{T}}
    \frac{\left( 1 - |\tilde{x}| \right) \sin{\left( K \left(\tilde{V}_R(x_i) + \tilde{V}_L(x_i) \right) \tilde{x} - \frac{\tilde{V}_R(x_i) - \tilde{V}_L(x_i) }{\pi \tilde{T}} (\tilde{y} - \frac{i \pi}{2})\right)}}
    {\left[ \left( \frac{1}{\pi \tilde{T}}\right)^2 
    \cosh{\left(-\pi \tilde{T} \tilde{x}+\tilde{y} \right)} 
     \cosh{\left(\pi \tilde{T} \tilde{x}+\tilde{y} \right)}  \right]^{mK}} \\
    \left< \hat{J}_{\Delta}^{(i)}(t) \right> = 
      \frac{i e \Delta^2}{2 \pi^2 u a } &
      \left( \frac{l}{a} \right)^{3-2m/K} 
    \int\displaylimits_{-1}^{1} d\tilde{x} \int\displaylimits_{-\infty-i\varepsilon+\frac{i \pi}{2}}^{\infty-i\varepsilon+\frac{i \pi}{2}} \frac{d \tilde{y}}{\pi \tilde{T}}
    \frac{\left( 1 - |\tilde{x}| \right) \sin{\left( \frac{1}{K}\left(\tilde{V}_R(x_i) - \tilde{V}_L(x_i) \right) \tilde{x} +  \frac{\tilde{V}_R(x_i) + \tilde{V}_L(x_i) }{\pi \tilde{T}} (\tilde{y} - \frac{i \pi}{2})\right)}}
    {\left[ \left( \frac{1}{\pi \tilde{T}}\right)^2 
    \cosh{\left(- \pi \tilde{T} \tilde{x}+\tilde{y} \right)} 
     \cosh{\left(\pi \tilde{T} \tilde{x}+\tilde{y} \right)}  \right]^{m/K}} \\
    \left< \hat{J}_{w,R/L}^{(i)}(t) \right> & = 
     \frac{i e |w|^2}{2 \pi^2 v_b a} \left( \frac{l}{a} \right)^{1-m\bar{K}} 
     \int\displaylimits_{-\infty-i\varepsilon+\frac{i \pi}{2}}^{\infty-i\varepsilon+\frac{i \pi}{2}} \frac{d \tilde{y}}{\pi \tilde{T}}
     \frac{\sin{\left( \frac{\tilde{V}_{R/L}(x_i)}{\pi \tilde{T}}(\tilde{y} - \frac{i\pi}{2})\right)}}{\left[ \left( \frac{1}{\pi \tilde{T}}\right)
    \cosh{\left(\tilde{y} \right)} 
     \right]^{m \bar{K}+1}}.
\end{align}
\end{subequations}
The first two equations have poles at $\tilde{y}=\pm \pi \tilde{T} \tilde{x}+i \pi (n+1/2)$, and the third at $\tilde{y}=i \pi (n+1/2)$, for any integer $n$. In particular, there are no poles between the line $\mathrm{Im}\left[\tilde{y}\right]=-i\varepsilon+i \pi/2$ and $\mathrm{Im}\left[\tilde{y}\right]=0$. So we can shift the integration contour back to the purely real axis. Additionally, all denominators are now even in both $\tilde{y}$ and $\tilde{x}$, allowing us to keep only even terms in the numerator. We thus obtain the full integral expressions 
\begin{subequations}
    \begin{multline}    
    \label{eq:ExpectationValues_Full_Backscattered}
        \left< \hat{J}_{\Gamma}^{(i)}(t) \right> = 
          \frac{e \Gamma^2}{2 \pi^2 u a } \left( \frac{l}{a} \right)^{3-2mK} 
        \sinh{\left( \frac{\tilde{V}_{R}(x_i)-\tilde{V}_{L}(x_i)}{2 \tilde{T}}\right)}
         \times  \\
        \int\displaylimits_{-1}^{1} d\tilde{x} \int\displaylimits_{-\infty}^{\infty} \frac{d \tilde{y}}{\pi \tilde{T}}
        \frac{\left( 1 - |\tilde{x}| \right) 
        \cos{\left( K \left(\tilde{V}_R(x_i) + \tilde{V}_L(x_i) \right) \tilde{x} \right)}
        \cos{\left(\frac{\tilde{V}_R(x_i) - \tilde{V}_L(x_i) }{\pi \tilde{T}} \tilde{y} \right)}}
        {\left[ \left( \frac{1}{\pi \tilde{T}}\right)^2 
        \cosh{\left(-\pi \tilde{T} \tilde{x}+\tilde{y} \right)} 
         \cosh{\left(\pi \tilde{T} \tilde{x}+\tilde{y} \right)}  \right]^{mK}}, 
    \end{multline} 
    \begin{multline}    
    \label{eq:ExpectationValues_Full_Andreev}
        \left< \hat{J}_{\Delta}^{(i)}(t) \right> = 
          \frac{e \Delta^2}{2 \pi^2 u a } 
          \left( \frac{l}{a} \right)^{3-2m/K} 
        \sinh{\left( \frac{\tilde{V}_{R}(x_i)+\tilde{V}_{L}(x_i)}{2 \tilde{T}}\right)}
          \times  \\
        \int\displaylimits_{-1}^{1} d\tilde{x} \int\displaylimits_{-\infty}^{\infty} \frac{d \tilde{y}}{\pi \tilde{T}}
        \frac{\left( 1 - |\tilde{x}| \right) 
        \cos{\left( \frac{1}{K}\left(\tilde{V}_R(x_i) - \tilde{V}_L(x_i) \right) \tilde{x}\right)}
        \cos{\left(  \frac{\tilde{V}_R(x_i) + \tilde{V}_L(x_i) }{\pi \tilde{T}} \tilde{y} \right)}}
        {\left[ \left( \frac{1}{\pi \tilde{T}}\right)^2 
        \cosh{\left(- \pi \tilde{T} \tilde{x}+\tilde{y} \right)} 
         \cosh{\left(\pi \tilde{T} \tilde{x}+\tilde{y} \right)}  \right]^{m/K}}, 
    \end{multline} 
    \begin{equation}    
    \label{eq:ExpectationValues_Full_Dissipated}
        \left< \hat{J}_{w,R/L}^{(i)}(t) \right>  = 
         \frac{ e |w|^2}{2 \pi^2 v_b a} \left( \frac{l}{a} \right)^{1-m\bar{K}} 
         \sinh{\left( \frac{\tilde{V}_{R/L}(x_i)}{2 \tilde{T}}\right)}
         \int\displaylimits_{-\infty}^{\infty} \frac{d \tilde{y}}{\pi \tilde{T}}
         \frac{\cos{\left( \frac{\tilde{V}_{R/L}(x_i)}{\pi \tilde{T}}\tilde{y} \right)}}{\left[ \left( \frac{1}{\pi \tilde{T}}\right)
        \cosh{\left(\tilde{y} \right)} 
         \right]^{m \bar{K}+1}}.
    \end{equation}
\end{subequations}

\subsubsection{Dissipated current, dissipation length}
\label{subsec:DissLength}
We first calculate the dissipating current given in Eq.~\eqref{eq:ExpectationValues_Full_Dissipated}. Simply performing the remaining integration and restoring the original variables, it gives
\begin{equation}
    \label{eq:Dissipation_current_exact}
    \left< \hat{J}_{w,R/L}^{(i)}(t) \right> = 
     \frac{ e |w|^2}{2 \pi^2 v_b a} \left( \frac{l}{a} \right)
     \left( \frac{2 \pi T a}{u} \right)^{m \bar{K}} \sinh{\left( \frac{e V_{R/L}(x_i)}{2 T}\right)} \mathcal{B} 
     \left( \frac{m \bar{K}}{2} + \frac{1}{2} + i\frac{e V_{R/L}(x_i)}{2 \pi T}, 
     \frac{m \bar{K}}{2} + \frac{1}{2} - i\frac{e V_{R/L}(x_i)}{2 \pi T}\right),
\end{equation}
where $\mathcal{B}(x,y)$ is the Euler Beta function. Focusing on the low voltage limit, $eV_{R/L}\ll T$, 
\begin{equation}
    \label{eq:Dissipation_current_low_voltage}
    \left< \hat{J}_{w,R/L}^{(i)}(t) \right> \approx 
     \frac{e |w|^2}{2 \pi u v_b } \left( \frac{l}{a} \right)
     \left( \frac{2 \pi T a}{u} \right)^{m \bar{K} - 1}  \mathcal{B} 
     \left( \frac{m \bar{K}}{2} + \frac{1}{2}, 
     \frac{m \bar{K}}{2} + \frac{1}{2} \right) e V_{R/L}(x_i) + O \left[\left( \frac{eV_{R/L}(x_i)}{T}\right)^3 \right].
\end{equation}
This allows us to define a characteristic dissipation length at which a fraction $0 \leq p \leq 1$ of all incoming current dissipates, $ \left< \hat{J}_{w,R/L}^{(i)}(t) \right> \approx p e^2 V_{R/L}(x_i)/hm$. We see from the expression above that this will be given by
\begin{equation}
    \label{eq:Dissipation_length}
    l_d^{-1} \approx \frac{1}{p} \frac{m |w|^2}{\hbar u v_b a}
     \left( \frac{2 \pi T a}{\hbar u} \right)^{m \bar{K} - 1}  \mathcal{B} 
     \left( \frac{m \bar{K}}{2} + \frac{1}{2}, 
     \frac{m \bar{K}}{2} + \frac{1}{2} \right),
\end{equation}
where we have restored $\hbar$. In particular, for non-interacting electrons ($m=\bar{K}=1$, $u = v$), the length at which all current dissipates ($p=1)$ is given by $l_d^{-1} \approx \frac{|w|^2}{\hbar v v_b a}$ and is completely temperature independent. This is essentially equivalent to a Fermi's golden rule transition rate between the edge and the dissipative bath. We henceforth use $p=1/e\approx 0.37$.

\subsubsection{Back-scattered and Andreev reflected currents}
\label{subsec:B_A_currents}
The back-scattered and Andreev reflected currents of Eq.~\eqref{eq:ExpectationValues_Full_Backscattered} and \eqref{eq:ExpectationValues_Full_Andreev} require more delicate care due to the integration over two separate indices. As we are interested only in the response at zero bias, we first approximate these expressions for $\tilde{V}_{R/L}(x_i) \ll 1$ and $\tilde{V}_{R/L}(x_i) \ll \tilde{T}$. The full integral expressions thus simplify to
\begin{subequations}
    \label{eq:ExpectationValues_LowVoltage}
    \begin{align}    
        \left< \hat{J}_{\Gamma}^{(i)}(t) \right> \approx &
          \frac{e \Gamma^2}{2 \pi^2 u a } \left( \frac{l}{a} \right)^{3-2mK} 
        \frac{\tilde{V}_{R}(x_i)-\tilde{V}_{L}(x_i)}{2 \tilde{T}}
        \int\displaylimits_{-1}^{1} d\tilde{x} \int\displaylimits_{-\infty}^{\infty} d \tilde{y}
        \frac{\left(\pi \tilde{T}\right)^{2mK-1} \left( 1 - |\tilde{x}| \right)}
        {\left[ 
        \cosh{\left(-\pi \tilde{T} \tilde{x}+\tilde{y} \right)} 
         \cosh{\left(\pi \tilde{T} \tilde{x}+\tilde{y} \right)}  \right]^{mK}}, \\
        \left< \hat{J}_{\Delta}^{(i)}(t) \right> \approx &
          \frac{e \Delta^2}{2 \pi^2 u a } 
          \left( \frac{l}{a} \right)^{3-2m/K} 
        \frac{\tilde{V}_{R}(x_i)+\tilde{V}_{L}(x_i)}{2 \tilde{T}}
        \int\displaylimits_{-1}^{1} d\tilde{x} \int\displaylimits_{-\infty}^{\infty} d \tilde{y}
        \frac{\left(\pi \tilde{T}\right)^{2m/K-1}\left( 1 - |\tilde{x}| \right)}
        {\left[ 
        \cosh{\left(- \pi \tilde{T} \tilde{x}+\tilde{y} \right)} 
        \cosh{\left(\pi \tilde{T} \tilde{x}+\tilde{y} \right)}  \right]^{m/K}}. 
    \end{align} 
\end{subequations}

For general parameters, the unitless number given by the integral can be calculated numerically. It is instructive however to focus on two particular limits where the integral can be calculated analytically as well. If $\pi \tilde{T} \ll 1$, then throughout the integration domain $\pi \tilde{T} \tilde{x} \ll 1$ as well. We thus replace $\cosh{\left(\pm \pi \tilde{T} \tilde{x}+\tilde{y}\right)}\approx \cosh{(\tilde{y})}$ in the denominators. The remaining integration over $\tilde{x}$ trivially gives 1, and integration over $\tilde{y}$ can again be described in terms of the Euler Beta function. Restoring all units,
\begin{subequations}
    \label{eq:ExpectationValues_LowVoltage_LowTemp}
    \begin{align}    
        \left< \hat{J}_{\Gamma}^{(i)}(t) \right>_{\tilde{T}\ll 1} \approx &
          \frac{e \Gamma^2}{2 \pi u^2  } \left( \frac{l}{a} \right)^{2}
          \left(\frac{2 \pi T a}{u}\right) ^{2mK - 2} \mathcal{B} \left( mK, mK\right)
        e\left(V_{R}(x_i)-V_{L}(x_i)\right) , \\
        \left< \hat{J}_{\Delta}^{(i)}(t) \right>_{\tilde{T}\ll 1} \approx &
          \frac{e \Delta^2}{2 \pi u^2 } 
          \left( \frac{l}{a} \right)^{2}  
          \left(\frac{2 \pi T a}{u}\right) ^{2m/K - 2} \mathcal{B} \left( \frac{m}{K}, \frac{m}{K}\right)
        e\left(V_{R}(x_i)+V_{L}(x_i)\right). 
    \end{align} 
\end{subequations}
Conversely, for the case $\pi \tilde{T} \gtrsim 1$, we see that the strongly divergent denominator in Eq.~\eqref{eq:ExpectationValues_LowVoltage} dictates that only values of $\tilde{x}< 1 / \pi \tilde{T} \lesssim 1$ contribute to the integral. We can thus approximate $\left( 1 - |\tilde{x}| \right)\approx 1$, and extend the integration domain to infinity, such that
\begin{align}    
        \left< \hat{J}_{\Gamma}^{(i)}(t) \right>_{\pi \tilde{T} \geq 1} \approx &
          \frac{e \Gamma^2}{2 \pi^2 u a } \left( \frac{l}{a} \right)^{3-2mK} 
        \frac{\tilde{V}_{R}(x_i)-\tilde{V}_{L}(x_i)}{2 \tilde{T}}
        \int\displaylimits_{-\infty}^{\infty} d\tilde{x} \int\displaylimits_{-\infty}^{\infty} d \tilde{y}
        \frac{\left(\pi \tilde{T}\right)^{2mK-1} }
        {\left[ 
        \cosh{\left(-\pi \tilde{T} \tilde{x}+\tilde{y} \right)} 
         \cosh{\left(\pi \tilde{T} \tilde{x}+\tilde{y} \right)}  \right]^{mK}}, \\
        \left< \hat{J}_{\Delta}^{(i)}(t) \right>_{\pi \tilde{T} \geq 1} \approx &
          \frac{e \Delta^2}{2 \pi^2 u a } 
          \left( \frac{l}{a} \right)^{3-2m/K} 
        \frac{\tilde{V}_{R}(x_i)+\tilde{V}_{L}(x_i)}{2 \tilde{T}}
        \int\displaylimits_{-\infty}^{\infty} d\tilde{x} \int\displaylimits_{-\infty}^{\infty} d \tilde{y}
        \frac{\left(\pi \tilde{T}\right)^{2m/K-1}}
        {\left[ 
        \cosh{\left(- \pi \tilde{T} \tilde{x}+\tilde{y} \right)} 
         \cosh{\left(\pi \tilde{T} \tilde{x}+\tilde{y} \right)}  \right]^{m/K}}. 
\end{align} 
The two integrals can now be decoupled by moving to light-cone variables $\pm \pi \tilde{T} \tilde{x}+y \equiv z_{\pm}$, with each integral contributing an Euler Beta function. We thus obtain
\begin{subequations}
    \label{eq:ExpectationValues_LowVoltage_HighTemp}
    \begin{align}    
        \left< \hat{J}_{\Gamma}^{(i)}(t) \right>_{\pi \tilde{T} \geq 1} \approx &
          \frac{e \Gamma^2}{4 \pi u^2  } \left( \frac{l}{a} \right)
          \left(\frac{2 \pi T a}{u}\right) ^{2mK - 3} \mathcal{B}^2 \left( \frac{mK}{2}, \frac{mK}{2}\right)
        e\left(V_{R}(x_i)-V_{L}(x_i)\right) , \\
        \left< \hat{J}_{\Delta}^{(i)}(t) \right>_{\pi \tilde{T} \geq 1} \approx &
          \frac{e \Delta^2}{4 \pi u^2 a } 
          \left( \frac{l}{a} \right)  
          \left(\frac{2 \pi T a}{u}\right) ^{2m/K - 3} \mathcal{B}^2 \left( \frac{m}{2K}, \frac{m}{2K}\right)
        e\left(V_{R}(x_i)+V_{L}(x_i)\right). 
    \end{align} 
\end{subequations}

We remind the reader that $\tilde{T}\equiv T l /u = l/l_T$, where $l_T \sim u/T$, is the thermal length. So the two cases $\tilde{T} \ll 1$ and $\tilde{T} \gtrsim 1$ correspond, respectively, to the cases $l \ll l_T$ and $l \gtrsim l_T$. In our configuration, motivated by the rather dominant dissipation observed in \cite{Ronen_2020}, we assume the main candidate to determine the length $l$ is the dissipation length, $l_d$. As such, the power-law dependencies of the currents on temperature depend on whether the system is in the $l_d \ll l_T$ regime or the $l_d\gtrsim l_T$ regime (see table \ref{table:Constants_A}).

Using the dissipation length we found in Eq.~\eqref{eq:Dissipation_length}, we see that $l_d \propto T ^{1-m\bar{K}}$.  As such, for $m \bar{K} > 2$, we are always in the regime $l_d > l_T$ for sufficiently low temperature. In particular, since $\bar{K} \geq 1$, this is always the case for $m \geq 3$. As such, we focus solely on this regime for Laughlin quantum Hall edges.

Conversely, for non-interacting integer states whose excitations are electrons, we found that $l_d$ is independent of $T$. As such, we cross over between the $l_d > l_T$ regime at high temperatures to $l_d < l_T$ at low temperatures.

\subsection{Results}
\label{subsec:Results}
Summing up all results, we find that, at the low voltage limit, we can always describe our expectation values as
\begin{subequations}
    \begin{align}
        \label{eq:ExpectationValues_Summary}
        \braket{\hat{J}_{\Gamma}^{(i)}} &= l_d \frac{e^2}{u^2 a} A_{\Gamma}(T) |\Gamma|^2 \left(V_R(x_i) - V_L(x_i) \right), \\
        \braket{\hat{J}_{\Delta}^{(i)}} &= l_d  \frac{e^2}{u^2 a} A_{\Delta}(T) |\Delta|^2 \left(V_R(x_i) + V_L(x_i) \right), \\
        \braket{\hat{J}_{w,R/L}^{(i)}} &= l_d  \frac{e^2}{u a} A_{w}(T) \alpha V_{R/L}(x_i),
    \end{align}
\end{subequations}
where $A_\Gamma(T)$, $A_\Delta(T)$ and $A_w(T)$ are unit-less and encode the temperature dependence. The explicit form of these temperature dependent coefficients is shown in Table~\ref{table:Constants_A}.
\renewcommand{\arraystretch}{1.8}

\begin{table}[h!]
\centering
    \begin{tabular}[c]{ |c|c|c| } 
     \hline
     $\quad$ & $l_d > l_T$ & $l_d < l_T$ \\ 
     \hline
     
     $A_\Gamma $ & $\frac{1}{4 \pi} 
              \left(\frac{2 \pi T a}{u}\right) ^{2mK - 3} \mathcal{B}^2 \left( \frac{mK}{2}, \frac{mK}{2}\right)
            $ & $\frac{1}{2 \pi} \left( \frac{l_d}{a} \right)
              \left(\frac{2 \pi T a}{u}\right) ^{2mK - 2} \mathcal{B} \left( mK, mK\right)
            $ \\[5pt]
     \hline
     $A_\Delta $ & $\frac{1}{4 \pi} 
              \left(\frac{2 \pi T a}{u}\right) ^{2m/K - 3} \mathcal{B}^2 \left( \frac{m}{2K}, \frac{m}{2K}\right)
            $ & $\frac{1}{2 \pi} \left( \frac{l_d}{a} \right)
              \left(\frac{2 \pi T a}{u}\right) ^{2m/K - 2} \mathcal{B} \left( m/K, m/K\right)
            $ \\[5pt]
     \hline
     $A_w$ & \multicolumn{2}{c|}{$ 
         \left( \frac{2 \pi T a}{u} \right)^{m \bar{K} - 1}  \mathcal{B} 
         \left( \frac{m \bar{K}}{2} + \frac{1}{2}, 
         \frac{m \bar{K}}{2} + \frac{1}{2} \right)$}  \\[5pt]
     \hline
    \end{tabular}
\caption{The temperature dependent coefficients $A_\Gamma$, $A_\Delta$ and $A_w$ as a function temperature $T$, inverse filling factor $m$ and Luttinger parameter $K$.}
\label{table:Constants_A}
\end{table}
For fractional quantum Hall states we focus solely on the regime $l_d > l_T$, whereas for non-interacting electrons we cross over between these two regimes at a critical temperature.

Now plugging these values into the Kirchhoff equation Eq.~\eqref{eq:Kirchhoff}, we obtain two classical, linear equations for the right- and left-edge voltages,
\begin{align*}
    \sigma_{xy} \left(V_R(x_{i+1})-V_R(x_i)\right) = &
    - l_d \frac{e^2}{u^2 a} A_{\Gamma} |\Gamma|^2 \left(V_R(x_i) - V_L(x_i) \right)
    - l_d \frac{e^2}{u^2 a} A_{\Delta} |\Delta|^2 \left(V_R(x_i) + V_L(x_i) \right)
    - l_d \frac{e^2}{u a} A_{w} \alpha V_{R}(x_i) \\
    \sigma_{xy} \left(-V_L(x_{i+1})+V_L(x_i)\right) = &
    + l_d \frac{e^2}{u^2 a} A_{\Gamma} |\Gamma|^2 \left(V_R(x_i) - V_L(x_i) \right)
    - l_d \frac{e^2}{u^2 a} A_{\Delta} |\Delta|^2 \left(V_R(x_i) + V_L(x_i) \right)
    - l_d \frac{e^2}{u a} A_{w} \alpha V_{L}(x_i). \nonumber
\end{align*}
Solving these equations using the two boundary conditions $V_R(0)=V$ and $V_R(L)=V_L(L)$, we obtain the solution 
\begin{align}
    \label{eq:Solution_Kirchhoff_General_L}
    \frac{dI}{dV} \bigg|_{V=0} = & \frac{e^2}{hm}
    \frac{ \left(1+l_d \lambda \right)^{\frac{L}{l_d}} 
    \left(A_{\Gamma} |\Gamma|^2 - A_{\Delta} |\Delta|^2 \right) + 
    \left(1-l_d \lambda \right)^{\frac{L}{l_d}} \left( A_{\Gamma} |\Gamma|^2 + A_{\Delta} |\Delta|^2 + u A_{w} \alpha + \lambda \right) }
    {\left(1-l_d \lambda \right)^{\frac{L}{l_d}}\left(A_{\Gamma} |\Gamma|^2 - A_{\Delta} |\Delta|^2 \right) +
    \left(1+l_d \lambda \right)^{\frac{L}{l_d}} \left( A_{\Gamma} |\Gamma|^2 + A_{\Delta} |\Delta|^2 + u A_{w} \alpha + \lambda \right)  },
\end{align}
where for convenience we define $(\lambda\frac{e}{hm})^2 \equiv \left( A_{\Gamma} |\Gamma|^2 + A_{\Delta} |\Delta|^2 + u A_{w} \alpha \right)^2 - \left( A_{\Gamma} |\Gamma|^2 - A_{\Delta} |\Delta|^2 \right)^2$. Taking the limit $L\rightarrow \infty$, this coincides with Eq.~\eqref{eq:Solution_Kirchhoff} of the main text.

\section{Dissipation at different filling fractions}
\label{app:fractions}

In this section we consider the CAR signature in systems with different filling fractions. We argue that if the dissipation is the most relevant perturbation, the system is \textbf{not} expected to exhibit a CAR signal which tends towards $-\sigma_{xy}$. If the dissipation is an irrelevant perturbation, the conductance will depend on the type of effective interactions between the edges. We then continue to examine the two filling fractions $\nu=2/5$ and $2/3$. In the experimental results of Ref.~\cite{Ronen_2020} these filling fractions show a qualitatively different CAR signal, wherein the $\nu=2/5$ case exhibits similar temperature depends to $\nu=1/3$ while $\nu=2/3$ exhibits similar behaviour to $\nu=1$. This is qualitatively consistent with the dissipation being irrelevant at $\nu=2/5$ and marginally relevant for $\nu=2/3$, as we explain below.  

The scaling behavior of the dissipation is determined by the properties of an electron tunneling operator, $O$, and its scaling dimension $\delta$. As per Appendix \ref{app:RG}, the dissipation can be treated as a non-local in time perturbation of the imaginary time action of the form
\begin{equation}
    \int dx\,d\tau_{1}\,d\tau_{2}\,\frac{-\alpha}{2\pi a^{2}}\frac{O^*(\tau_{1},x)O(\tau_{2},x)}{\tau_{1}-\tau_{2}}.
\end{equation}
The RG flow equation of $\alpha$ to first order in $\alpha$ is
\begin{equation}
\alpha^{-1}d\alpha/d\ell = 2-2\delta,
\end{equation}
and therefore dissipation is an irrelevant perturbation if $\delta>1$ and is relevant if $\delta<1$. Considering now two opposite edges across a finger, we saw in the main text that the scaling dimension $\delta$ depends on the inter-edge electronic interactions.

In the case that the dissipation is the most relevant perturbation, the system will renormalize at low temperatures to vanishing inter-edge interactions, just as in the integer filling case, similar to Eq.~\eqref{eq:RG eqns}, in which a large $\alpha$ drives $K$ towards $K=1$. The scaling dimension $\delta$ is then determined by the theory describing a single edge. The analysis of the scaling dimension of the dissipation using a single edge is then consistent if $\delta<1$, corresponding to the dissipation being a relevant perturbation.

In the contrary case where the dissipation is irrelevant, attractive inter-edge effective interactions can, in general, lead to the superconducting term being less irrelevant than the dissipation (for sufficiently effective attractive interactions, the superconducting term can be relevant). Consequently, this leads to an increase in the CAR signal at low temperatures, just as in our analysis of Laughlin edge states.

We now specialize to the cases of filling fractions $\nu=2/5,2/3$.  The topological classification of the FQH edge state will determine whether the dissipation term is relevant or not. For abelian FQH states this classification is given by the $K$-matrix and charge vector (see for example Ch.~7 in  Ref.~\cite{wen_quantum_2004}). For both the $\nu=2/3$ case and the $\nu = 2/5$ case we consider two inequivalent topological states:
\begin{itemize}
    \item Consider the case of filling $\nu=2/m$, with the double layer topological state of
\begin{equation}
    K=\begin{pmatrix}
        m & 0\\
        0 & m
    \end{pmatrix},\quad
    q = 
    \begin{pmatrix}
        1\\
        1
    \end{pmatrix}.
    \end{equation}
    
    A general operator corresponding to an integer number of electrons tunneling into the edge is given by 
    $O=e^{i(n_1,\,n_2)K(\phi_1,\,\phi_2)^t}$ with $n_1,n_2$ integers, and adds charge $Q=(n_1,\,n_2) \cdot q=n_1+n_2$. We assume a generic fixed-point Hamiltonian $H=\int dx\,\left\{v_1 (\partial_x \phi_1)^2+v_2(\partial_x\phi_2)^2+2v_{12}\partial_x\phi_1\partial_x\phi_2\right\}$ with $v_{1}v_{2}>v_{12}^2$ and $v_1,v_2>0$, so that its spectrum is bounded from below. Since in this topological state all fields are chiral left-movers, then the operator $O$ has scaling dimension 
    \begin{equation}
    \delta = (n_1,\,n_2)K(n_1, n_2)^t /2= m(n_1^2 + n_2^2)/2.
    \end{equation}
    For an electron tunneling operator, $Q=1$, the minimal scaling dimensions is attained for $(n_1,\,n_2)=(1,\,0)\text{ or }(0,\,1)$ with  $\delta=m/2$. Thus, electron dissipation is irrelevant at filling $\nu=2/m$ in this topological state.

    \item At filling $\nu=2/5$ we consider the topological edge state with $K$-matrix and charge vector 
\begin{equation}
    K=\begin{pmatrix}
        3 & 2\\
        2 & 3
    \end{pmatrix},\quad
    q = 
    \begin{pmatrix}
        1\\
        1
    \end{pmatrix}.
\end{equation}
This topological state has two left-moving chiral fields, and thus can be analyzed in a similar fashion to the  $\nu=2/m$ case. The electron tunneling operators have the minimal scaling dimension of $\delta=3/2$, which is attained by the two operators $O=e^{i(n_1,\,n_2)K(\phi_1,\,\phi_2)^t}$ with $(n_1,\,n_2)=(1,\,0)$ and $(0,\,1)$. Consequently, the dissipation is irrelevant in this topological state.

\item For the standard single layer FQH edge $\nu=2/3$ with $K$-matrix and charge vector
\begin{equation}
    K=\begin{pmatrix}
        1 & 0\\
        0 & -3
    \end{pmatrix},\quad
    q = 
    \begin{pmatrix}
        1\\
        1
    \end{pmatrix}
\end{equation}
some more care is needed. We argue that the dissipation is relevant in most of experimentally accessible systems. In this topological state the scaling dimension of an electron tunneling operator depends on the fixed-point Hamiltonian of the system, which takes the general form $H=\int dx\,\{v_1 (\partial_x \phi_1)^2+3v_2(\partial_x\phi_2)^2+6v_{12}\partial_x\phi_1\partial_x\phi_2\}$, where $v_{1}v_{2}>3v_{12}^{2}$ and $v_1,v_2>0$. 
In the presence of charge conserving disorder the stable fixed-point is obtained at $R \equiv 2v_{12}/(v_1+v_2)=1/2$ (see Ref.~\cite{Kane_1994}). In this fixed-point the electron tunneling operators with the smallest scaling dimension are $e^{i\phi_1}$ and $e^{2i\phi_1+3i\phi_2}$, with $\delta=1$. Thus, to lowest order the dissipation term  is marginal. 

Going beyond leading order, the dissipation is likely to be marginally relevant due to a variety of effects. First, the dimensionless parameter $\alpha/v$, with $v$ some velocity scale of the fixed-point $H$, grows due to renormalization of the velocity to smaller values [as in Eq.~\eqref{eq:RG eqns} of the main text]. Furthermore, second order $\alpha^2$ corrections result in an RG flow equation of the form $d\alpha/d\ell = \alpha^2/[(v_1+v_2) C]$, where $C=\frac{\pi /4}{2\ln(2+\sqrt{3})-\sqrt{3}}$ (we omit the derivation for brevity).  Finally, non-charge conserving disorder can introduce slight deviations from the $R=1/2$ fixed-point, due to proximity to a SC. This causes one of the tunneling operators, $e^{i\phi_1}$ or $e^{2i\phi_1+3i\phi_2}$, to have $\delta<1$, and thus  the dissipation becomes relevant. More specifically, in the region $-1/2<R<1/2$ the tunneling operator $e^{i\phi_1}$ has $\delta<1$, and in the region $1/2<R<15/26$ the tunneling $e^{2i\phi_1+3i\phi_2}$ has $\delta<1$.
\end{itemize}
\end{widetext}

\end{document}